\documentclass[usenatbib,onecolumn]{mn2e}
\usepackage{graphicx}
\usepackage{amsmath}
\usepackage[usenames,dvipsnames]{color}
\newcommand{\hmpc}{ \, h \rm{ Mpc}^{-1}}
\def\lsim{~\rlap{$<$}{\lower 1.0ex\hbox{$\sim$}}}
\def\gsim{~\rlap{$>$}{\lower 1.0ex\hbox{$\sim$}}}

\begin{document}

\title[Bias of weighted DM halos]{The bias of  weighted dark matter halos from peak theory}

\author[Verde et al.]{Licia Verde$^{1,2,3,5}$,\thanks{email: liciaverde@icc.ub.edu} Raul Jimenez$^{1,2,4,5}$, Fergus Simpson$^{2}$, Luis Alvarez-Gaume$^5$, \newauthor Alan Heavens$^6$ and Sabino Matarrese$^{7,8,9}$\\
$^1$ ICREA (Instituci\'o Catalana de Recerca i Estudis Avan\c{c}at).\\
$^{2}$  Instituto de Ciencias del Cosmos, University of Barcelona, UB-IEEC, Marti i Franques 1, E08028, Barcelona, Spain.\\
$^3$ Institute of Theoretical Astrophysics, University of Oslo, 0315 Oslo, Norway. \\
$^4$ Institute for Applied Computational Science, Harvard University, MA 02138, USA.\\
$^5$ Theory Group, Physics Department, CERN, CH-1211, Geneva 23, Switzerland.\\
$^6$ Imperial Centre for Inference and Cosmology, Imperial College, Blackett Laboratory, Prince Consort Road, London SW7 2AZ U.K. \\
$^7$ Dipartimento di Fisica e Astronomia G. Galilei, Universit\'a degli Studi di Padova, I-35131 Padova, Italy. \\
$^8$ INFN, Sezione di Padova, I-35131 Padova, Italy. \\
$^9$Gran Sasso Science Institute (INFN), viale F. Crispi 7, 67100 L'Aquila, Italy}

\maketitle

\begin{abstract}
We give an analytical form for the weighted correlation function of peaks in a Gaussian random field. In a cosmological context, this approach strictly describes the {\it formation bias} and is the main result here. Nevertheless, we show its validity and applicability to the {\em evolved} cosmological density field and halo field, using  Gaussian random field realisations and dark matter N-body numerical simulations. Using this result from peak theory we compute the bias of peaks (and  dark matter halos) and show that it reproduces results from the  simulations at the ${\mathcal O}(10\%)$ level. Our analytical formula for the bias predicts a scale-dependent bias with two characteristics: a broad band shape which, however, is most affected by the choice of weighting scheme and evolution bias, and a more robust, narrow  feature localised  at the BAO scale, an effect that is confirmed  in simulations. This scale-dependent bias smooths the BAO feature but, conveniently, does not move it. We provide a simple analytic formula to describe this effect. We envision that our analytic solution will be of use for galaxy surveys that exploit galaxy clustering. 
\end{abstract}


\section{Introduction}
One of the biggest challenges of large-scale structure surveys is to infer the properties of the dark matter density field from observables such as galaxies or clusters.
 Galaxies, and the dark matter halos they inhabit, are not  perfect  tracers of the underlying dark matter distribution, but it is the statistical properties of the dark matter distribution that are most robustly predicted by theory. Modelling of the clustering properties of the dark matter halos, or more precisely, modelling of the halo bias, has received recently renewed attention  (e.g., \citet{Baldauf/etal:2013, Paranjape/etal:2013, Desjacques:2013,Castorina/Sheth:2013,Musso/Paranjape/Sheth:2012,Elia/etal:2011,Elia/etal:2012}) but pioneering work dates back to the  1970s-1980s (e.g., \cite{Doroshkevich:1970a,Doroshkevich:1970b,Kaiser84,JensenSzalay86}) as we will review below.
Modelling the  halo bias is particularly interesting for several reasons: the clustering of halos is driven only by gravity and thus in principle is completely specified by the initial conditions; it is virtually unaffected  --at least on scales  significantly larger than  the size of the halos, ${\cal O}(Mpc)$-- by  poorly known baryonic physics and  physics of galaxy formation (which instead drive  galaxy bias). It is also a crucial intermediate step to a full modelling of galaxy bias, if, for example, a halo occupation distribution model is used to describe how galaxies populate halos (e.g., \cite{Seljak:2000, Peacock/Smith:2000, CooraySheth02}).  In principle a galaxy  survey could be engineered so that the selected  galaxies trace dark matter halos,   for example by targeting bright luminous red galaxies which  are typically central halo galaxies (e.g., \cite{Mandelbaum/etal:2006}). In practice, successful attempts have been made to reconstruct the halo density field from a real galaxy survey \citep{reid/spergel:2008, reid/spergel/bode:2009, reid/etal:2009}. Clusters of galaxies also are believed to trace the  spatial distribution of (high mass) dark matter halos, and many forthcoming surveys   promise to provide cluster correlation properties.
It is well known that bias depends on halo properties, and in general halo bias is expected to be complicated, non-linear, non-local and scale-dependent. However, an accurate understanding of its behaviour is crucial to extract precise cosmological information from large scale structure clustering.  For example, the shape and amplitude of the matter power spectrum, or equivalently its correlation function, are sensitive to cosmological parameters such as neutrino masses. Also, the primordial power spectrum slope and shape, and the precise location of the Baryon Acoustic Oscillation (BAO) can provide a direct probe of the Universe's expansion history. Future galaxy surveys will probe an appreciable fraction of the observable universe, reducing the statistical errors on these quantities and making the scale-dependence and non-linearity of halo bias a source of systematic error that cannot be ignored.

 The halo bias could be studied and modelled, in principle,  solely via N-body numerical simulations (e.g., \cite{Seljak/Warren:2004, Paranjape/etal:2013,Elia/etal:2012}). However in practice the calculations needed to obtain the desired accuracy and error estimation far exceed the amount of CPU time available (see e.g., \cite{Dodelson/Schneider:2013,Morrison/Schneider:2013}). Having an analytic expression would be highly valuable: it could be used  for example to model the halo bias scale dependence and/or the cosmology dependence, thus having to rely on  N-body simulations only for calibrating and validating the analytic expressions.
 Further, it is always much more insightful to obtain a physical understanding of phenomena, such as the clustering of dark matter halos with respect to the dark matter field. In fact while N-body simulations have confirmed the non-linearity, non-locality, stochasticity and scale-dependence of halo bias, the origin of these effects remain unclear (see e.g., \citet{Porciani:2013}  and references therein).
 
 In this paper we  show how, using peak theory, we can derive an analytic expression for  the correlation properties of the dark matter peaks which, we argue, can largely be identified with  dark matter halos; our expression  depends on  the power spectrum of the dark matter field. This approach does not  model the bias itself, however it provides a description of the observable quantity (the correlation properties of peaks/halos)  from which a ``bias" can be obtained from e.g.,  the ratio of the relevant power spectra.
The  rest of the paper is organised as follows: In \S \ref{sec:review-and-approach} we review the current knowledge  on  halo bias and present the  aim and goals of our approach. In \S \ref{sec:method} we present our derivation and the  analytic expression for the halo bias. We also discuss the unavoidable approximations involved and their possible limitations. Sec. \ref{sec:validation} validates the approximations made and evaluates the performance of the formula comparing with simulations and in \S \ref{sec:BAO}  we present the consequences and possible applications of our findings especially for Baryon Acoustic Oscillations studies. Finally we conclude in \S \ref{Conclusions}.

\section{Review of current state of affairs and  our approach}
\label{sec:review-and-approach}
Several different approaches have been used in the literature  to model or understand  halo bias.
The common denominator is to define the bias as a function  relating the dark matter overdensity field $\delta_m(x)$ to the halo overdensity field $\delta_h(x)$, $\delta_h(x)=f(\delta_m(x))$. This relation is then often expanded around small $\delta_m(x)$ where the expansion coefficients are the bias parameters (see e.g., \cite{HMV,McDonald:2006, 2009JCAP...08..020M} and refs therein). In this framework,  the approaches describing the abundance of collapsed objects  can be extended to describe the halo bias (via a peak-background split argument e.g., \cite{ST99,Schmidt/Jeong/Desjacques:2013} and refs therein).  Analytical descriptions of the halo mass function aim to characterize the location of collapse {\it in the initial conditions}. Two parallel approaches have been traditionally investigated: the peaks formalism (e.g., \cite{Peacock/Heavens:1985,BBKS, Cole:1991, Lumsden/etal:1989,Lumsden/etal:1990}) and the excursion set (e.g., \cite{PS74,Bond/etal:1991,ST99,Sheth/Mo/Tormen:2001,Maggiore/Riotto:2010, Musso/Paranjape/Sheth:2012,Paranjape/Sheth:2012b} and references therein). While the peak formalism treats density peaks as special sites for halo formation the excursion set treats all locations in the initial conditions on the same footing.  The ``peak-patch" approach  \citep{Bond/Myers:1996} tried to unify the two, a unification that  was recently achieved analytically (\cite{Paranjape/Sheth:2012a,Paranjape/Sheth/Desjacques:2013}, denoted {\it excursion set of peaks}) by making simple approximations that are supported by N-Body experiments.  All approaches  rely on the statistical properties of the initial conditions to predict final halo properties. In particular  the study of {\it Lagrangian bias} is considered an important step in understanding bias, with the hope that the mapping between Lagrangian and Eulerian space   is simple \citep{MoWhite96,Jing:1999, CLMP98}.  In other words halo bias could be split into two parts: a {\it formation bias},  which we identify with the Lagrangian bias and which represents halos being born more strongly clustered than the matter, and an {\it evolution bias}.  The latter is a mixture of complex effects from effects such as nonlinear clustering, movement of matter, and merging. The formation bias is expected to dominate with  evolution bias adding a  small correction.

While the bias predicted by the peak background split does not seem to match N-body simulations results \citep{Manera/Sheth/Scoccimarro:2010}, it has recently been confirmed that  a large fraction of haloes  originate from initial density peaks \citep{Ludlow/Porciani:2011}. In particular the excursion set of peaks approach gives a description of halo bias accurate at the 10\% level, which could be improved by a remapping of the assignment of masses \citep{Hahn/Paranjape:2014}.  This finding then offers supports to the approach that 
   Lagrangian bias can be well described by studying the peaks of the initial density field. As initial conditions are believed to be Gaussian (and supported by microwave background measurements, \cite{PlanckXXIV}) it is therefore interesting to investigate the properties of peaks of Gaussian random fields. The relation  between overdensity of peaks and overdensity of matter $\delta_{\rm pk}=f(\delta_m)$ might be still complicated.  In fact, despite this simplification, halo bias  is  still complicated: non-local, non-linear, stochastic, and  the expansion  of the $\delta_h(x)=f(\delta_m(x)) $ relation  needs many coefficients (bias parameters) e.g., \cite{2001ApJ...558..520Y,Seljak/Warren:2004,Baldauf/etal:2013, Ludlow/Porciani:2011, Pollack/etal:2013,Castorina/Sheth:2013,Sheth/Chan/Scoccimarro:2013}. On the other hand, 
 the clustering properties of peaks and especially their two-point function  can have a direct relation to the  matter one. In one dimension the correlation function of maxima of a Gaussian field has been shown to be directly related to  the correlation function of the field itself \citep{Adler:1981,Peacock/Heavens:1985,BBKS}. In more than two dimensions, however, the (N-point) correlation function of maxima of a Gaussian field does not have, so far, relatively simple, closed expressions. The usual approach is to find closed-form solutions which are evaluated numerically (see e.g. \citet{heavens/sheth:99} for two dimensions on the sky, for small angles, and \citet{HeavensGupta} for all-sky). This is a long standing open problem starting from the first attempt by \citet{Otto/Politzer/Wise:1986,Otto/Politzer/Wise:1986err}. 
 
Directly computing the peak correlation function  has the advantage that it does not rely on expanding the bias relation. \citet{Desjacques:2008} and  \citet{Desjacques/Crocce/etal:2010} took a similar approach.   \citet{Desjacques:2008}  computes the large scale expression  for the correlation function of peaks and finds that spatial derivatives of the linear density correlation functions have important effects on the clustering of peaks. We will return to this finding in this work.   \citet{Desjacques/Crocce/etal:2010}  derive an expression of the two-point correlation function of initial density peaks but in a perturbative  way to second order in the density.   Their findings include a  scale-dependent bias even on relatively large scales corresponding to the BAO feature and enhancement of the feature. They further explore the  effects and scale-dependence of evolution bias.  Contrary to \citet{Desjacques/Crocce/etal:2010} we find that the  peaks scale-dependent bias introduces a reduction of the BAO feature rather than an enhancement. Here we take the approach that rather than describing the bias function,  a quantity of  interest is the correlation function or power spectrum of halos.  By {\it bias} we refer to the square root of the ratio of the correlations (or power spectra) of the tracers and the field. Further, we concentrate on formation bias by using the clustering of peaks of the initial Gaussian field as a proxy for the clustering of halos, and this is the main result that is presented here.  The effects of evolution bias are  investigated  and quantified as a second step.   

In summary in this paper we present an analytic solution to the (N-point) correlation function of extrema for any dimensions and then focus on the  two point function in three dimensions, which is of most practical relevance.  Since most extrema  above practically interesting thresholds ($>2\sigma$) are peaks  we  find that can identify the clustering of extrema with the clustering of peaks. We investigate the effects of  our unavoidable approximations  on a suite of Gaussian realizations and then show how  this formula performs  for peaks and halos of an {\it evolved} density field by comparing to N-body simulations. 

\section{Method}
\label{sec:method}

Let us start from Eq.~(3) of \citet{Otto/Politzer/Wise:1986}  expressing the joint probability of finding $N$ peaks of a field $\phi({\bf r})$ at positions ${\bf r}_i$, $i=1,..,N$,  and above a threshold, which we report here:

\begin{equation}
P({\bf r}_1,\ldots ,{\bf r}_N)=\int [d\phi({\bf r})] P[\phi({\bf r})]  \prod_{j=1}^N\left[ \int dw_{(j)} |\det w_{(j)}|\delta^3(\nabla\phi({\bf r}_j)) \delta^6(\nabla \nabla \phi(r_j)-w_{(j)})\theta(\phi({\bf r}_j)-t)\right]\,.
\label{eq:path}
\end{equation}

\citet{Otto/Politzer/Wise:1986} use the  extremely powerful path integral approach to describe  cosmological Gaussian random fields; this technique has been used with remarkable success in cosmology.
On the assumption that galaxies and clusters of galaxies, or dark matter halos, occur at local maxima of the field $\phi({\bf r})$ that are above the threshold $t$, this expression gives  the probability of finding $N$ objects at locations ${\bf r}_1,...,{\bf r}_N$. In this equation $P[\phi]$ is the Gaussian probability distribution function,  $\theta$ denotes the Heaviside step function and $w_{(j)}$ in our adopted notation and in three spatial dimensions is the symmetric $3\times3$ matrix of the second derivatives of $\phi$ at position ${\bf r}_j$. In Eq.~(\ref{eq:path}) the integration on $dw_{(j)}$ has to be extended only over negative definite eigenvalues, in order to identify local maxima. 

The correlation function is then given by
\begin{equation}
1+\xi_{1,2} =  P({\bf r}_1,\ldots ,{\bf r}_N)/P^N({\bf r}_1)\,.
\end{equation}
The expression for $P({\bf r}_1)$ was presented in \citet{BBKS}.

We then make the approximation that halos correspond to local maxima of the linearly extrapolated initial Gaussian field. 
As such, we only have to work out the properties of peaks in Gaussian fields, which have been extensively studied in the literature \citep{Rice:1944,Rice:1945,Adler:1981, BBKS, Peacock/Heavens:1985,Kaiser84,JensenSzalay86}.

This is therefore equivalent to computing the Lagrangian or formation bias  (see \S \ref{sec:review-and-approach}) except that here we will define bias as the square-root of the ratio of the correlation function. By making this approximation our description will not include non-linear clustering of halos,  the fact that halos might move from initial positions, merge or that  halos at low redshift  might not correspond to initial peaks. We expect these corrections to be small at least for relatively massive halos (e.g., \cite{Ludlow/Porciani:2011}) and we will further quantify them in \S \ref{sec:validation}.   Our result will only include effects induced by the highly non-linear transformation of a Gaussian field which is applied when selecting only the location of peaks above a threshold.
Thus in  Eq.~(\ref{eq:path})  the first integral is a (Gaussian)  path integral, with $\phi$ being a Gaussian random field and thus 
\begin{equation}
P[\phi({\bf r})] \propto \exp[-\frac{1}{2}(\phi, K,\phi)]
\end{equation}
where we have used the short-hand notation for bivariate forms and operators.  The $\propto$ sign suppresses the normalization factor $(\det K/(2\pi)^M)^{1/2}=\{\int [d\phi] \exp[-1/2(\phi,K,\phi)]\}^{-1}$ which will cancel out later. $K$ is defined by (see  e.g.,\citet{Politzer/Wise:1984,Verdepdf2013} 
\begin{equation}
\int d^3y K(|{\bf x}-{\bf y}|)\xi(|{\bf y}-{\bf z}|)=\delta(|{\bf x}-{\bf z}|)\,.
\end{equation}

In order to perform the integrals in Eq.~(\ref{eq:path}) we note that while mathematically correct, the second-derivative condition to determine the maxima  significantly complicates the calculations and might be unnecessarily strong. 
Let us notice that if one integrates over all possible values of $w$ one would obtain extrema. However it is not a bad approximation to assume that for thresholds $t$ not too low, almost  all extrema will  be (local) maxima (we demonstrate this point in \S ~\ref{sec:validation}). More in general the approximation of neglecting the second derivative condition and its integrals will be very good for $t> {\rm few}$ ${\rm rms}$ \citep{Adler:1981,BBKS}, especially since we are interested in correlation functions, which are statistical quantities averaged over all pairs  as a function of their distance. This consideration allows us to neglect the Dirac delta function of the second derivatives. 

The term $|\det w_{(j)}|$ however remains. In fact if we want the correlation function of peaks (or extrema, or critical points) it means that each peak (or critical point) will count $1$ regardless of the spatial volume it occupies. If we wanted the correlation of regions above the  threshold then $|\det w_{(j)}|$ would not be there (e.g., \cite{Kaiser84,JensenSzalay86} ). 
 In fact $|\det w_{(j)}|$ is the Jacobian of the transformation  from $\nabla \phi $ to ${\bf r}$.  It is easier to see this in one dimension: $d\nabla \phi =(d\nabla \phi/dr) dr$ but if we want a number density the integral must be divided by the volume. So the $dr$  at the end cancels out but $d \nabla \phi$ (or $|\det w|$) remains.  
 
Unfortunately  $|\det w|$ makes the expressions not analytic. However it is clear that  it acts as a  weight and thus in principle it could be compensated by suitably weighting the data. We  start addressing this in the appendix   where we propose a  relatively straightforward way to compute  a proxy for this weight   from the data, i.e. in real galaxy surveys, thus making it possible to reintroduce this factor from the data themselves. Therefore  let us drop it from the equation  for the moment and keep in mind that it will have to be  reintroduced  later on as a weighting scheme on the observations. This manipulation will allow us to obtain analytic expressions for the correlation function.
 
Let us therefore work with the simplified expression
\begin{equation}
P({\bf r}_1,\ldots ,{\bf r}_N)=\int [d\phi({\bf r})] P[\phi({\bf r})] \prod_{j=1}^N \int_{m_j=t}^{\infty}\delta(\phi({\bf r}_j)-m_j) \delta^3(\nabla \phi({\bf r}_j) )dm_j\,.
\end{equation}

We will leave the $\int dm$ to be  performed at the end as it is similar to an error function and work with its integrand which we now denote $P'$ so that 
 \begin{equation}
 P({\bf r}_1,\ldots,{\bf r}_N)=\int_{t}^{\infty} dm_1...dm_N P'({\bf r}_1,\ldots,{\bf r}_N,m_1,..,m_N)\,.
 \end{equation}
Using the Fourier representation of the Dirac delta
\begin{equation}
\delta(s)=\int \frac{d\alpha}{2\pi} \exp[is\alpha]
\end{equation}
we obtain
\begin{eqnarray}
P'({\bf r}_1,\ldots,{\bf r}_N,m_1,..,m_N)\!\!\!\!&\!\!\!=\!\!\!&\!\!\!\int [d\phi({\bf r})] P[\phi({\bf r})] \prod_{j=1}^N\delta(\phi({\bf r}_j)-m_j) \delta^3(\nabla \phi({\bf r}_j) ) \nonumber \\
&=&\int d\alpha_1 ... d\alpha_N d\beta_{1,1}, d\beta_{1,2}.... d\beta_{N,d} %\\
%&&\!\!\!\!\!
 \exp\left[-i\sum_{j=1}^N\alpha_j m_j\right] 
\exp[i(J,\phi)] \exp[-\frac{1}{2}(\phi, K, \phi)] \nonumber
\end{eqnarray}
 where $d$ denotes the number of  spatial dimensions (1, 2 or 3) and we have defined the ``source" functional
 \begin{equation}
 J({\bf r})=\sum_{jp}(\alpha_j+\beta_{jp}\nabla_p)\delta({\bf r}-{\bf r}_j)\,.
 \end{equation} 
 The index $p$ (and $q$ later) runs over the number of dimensions and indicates a component.  Other labels ($i,j$ etc) label the points.
 
 The path integral in $\phi$ can be readily performed, leading to
 \begin{equation}
P'({\bf r}_1,\ldots ,{\bf r}_N,\vec{m})= \int d\{\alpha_j\} d\{\beta_{jp}\} \exp\left[-i\sum_{j=1}^N\alpha_j m_j\right] \exp\left[-\frac{1}{2}(J,K^{-1},J)\right]
 \end{equation}
 where we have used the shorthand notation of $d\{\alpha_j\}= d\alpha_1....d\alpha_N$ and the normalization factors nicely cancel out.
 
 In the Gaussian case $K^{-1}$ can be interpreted as the correlation function $\xi$  of the field $\phi$, which depends only on the distance between any two points not their location or orientation $\xi({\bf x}_j,{\bf x}_l)\equiv\xi(|{\bf x}_j-{\bf x}_l|)=\xi_{jl}=\xi(r_{jl})$.
 
 Let us now study  the $(J, K^{-1},J)$ expression,
 \begin{equation}
 (J, K^{-1},J) = \sum_{jl}\alpha_j \xi_{jl}\alpha_l  \sum_{jlpq} \beta_{jp}\beta_{lq}\nabla_{jp}\nabla_{lq}\xi_{j,l} + \sum_{jlp}\alpha_j\beta_{lp}\nabla_{lp}\xi_{jl}+ \sum_{jlp}\alpha_l\beta_{jp}\nabla_{jp}\xi_{jl}\,.
 \end{equation}
 Let us define a vector
 $\lambda={\alpha_1..\alpha_N, \beta_{1x}, \beta_{1y}, \beta_{1z},...,\beta_{Nx},\beta_{Ny},\beta_{Nz}}$ for three dimensions.
 Then the above quadratic form can be seen as a matrix operation $\lambda^T X\lambda$ where $X$ is a block matrix
 \begin{equation}
 \begin{matrix}
 X
 \end{matrix}
 =
\begin{pmatrix}
 \xi  & \nabla \xi \\
 \nabla \xi^T & \nabla \nabla \xi
 \end{pmatrix}
\end{equation}
and $\xi$ is the symmetric matrix given by the correlation function; $\nabla \xi$ has $N$ rows and $N\times d$ columns. The ordering we have chosen is so that we have $N$  $N\times d$ blocks. The evaluation of the block elements requires some caution as we will see below.  Finally $\nabla\nabla\xi$ is a matrix made by $N\times N$ sub blocks of size $d\times d$  whose elements are $\partial ^2 \xi({\bf x}_j,{\bf x}_l)/\partial r_{jp}\partial r_{lq}$. Let us define $r_{ij}=\sqrt{\sum_p (r_{ip}-r_{jp})^2}$ but ${\bf r}_{ij}={\bf r}_i-{\bf r}_j$.
Given the properties of $\xi$ we can write down expressions for these terms:
\begin{equation}
\frac{\partial \xi ({\bf r}_j,{\bf r}_l)}{\partial r_{lp}}=\frac{d\xi(r_{jl})}{dr}\frac{\partial r_{jl}}{\partial r_{lp}}=-\xi'(r_{jl})\frac{({\bf r}_j-{\bf r}_l)_p}{r_{jl}}\label{eq:deriv1}
\end{equation}
\begin{equation}
\frac{\partial \xi _{jl}}{\partial r_{jp}}=\frac{d\xi(r_{jl})}{dr}\frac{\partial r_{jl}}{\partial r_{jp}}= \xi'(r_{jl})\frac{({\bf r}_j-{\bf r}_l)_p}{r_{jl}}\,.
\label{eq:deriv2}
\end{equation}
Despite $\xi$ depending only on the distance,  Eq.\ref{eq:deriv1} and \ref{eq:deriv2}  are different, as a  small change in $r_j$ affects the  selected shape of the N-point function differently from that of a small change in $r_l$. In addition,
\begin{eqnarray}
\frac{\partial ^2 \xi_{jl}}{\partial r_{jp}\partial r_{lq}}&=&\left[\frac{\xi' (r_{jl})}{r_{jl}}-\xi''(r_{jl})\right]\frac{({\bf r}_j-{\bf r}_l)_p({\bf r}_j-{\bf r}_l)_q}{r^2_{jl}}-\delta_{pq}\frac{\xi'(r_{jl})}{r_{jl}} %\nonumber \\
%&=&
= {\cal H}(r_{jl}) (r_{jp}-r_{lp})(r_{jq}-r_{lq})-\delta^K_{pq}\frac{\xi'(r_{jl})}{r_{jl}}\,.
\label{eq:seconderxi}
\end{eqnarray}

It is useful now to express the above elements in terms of moments of the field. 
 \begin{eqnarray}
 \xi_{ij}&=&\frac{1}{4\pi^2 r}\int k P(k) \sin(kr) dk=\frac{1}{4\pi^2 r}\sum_{n=0}^{\infty}\frac{(-1)^n}{(2n+1)!}r^{2n+1}\int dk k^{2 n+2} P(k) %\nonumber \\
 %&& 
 =\sum_n \frac{(-1)^n}{(2n+1)!}r^{2n}\sigma_{2n+2}
 \end{eqnarray}
 where 
 \begin{equation}
 \sigma_n\equiv\frac{1}{4\pi^2}\int dk k^n P(k).
 \end{equation}
 This requires an intrinsic smoothing scale to be discussed later. Nevertheless we can expand the correlation function for small separations as:
 \begin{equation}
 \xi(r)=\sigma_2-\frac{\sigma_4}{3!}r^2+\frac{\sigma_6}{5!}r^4...
 \end{equation}
 giving the  behaviour for $\nabla \nabla \xi$  when $j=l$, $r\rightarrow 0$:
 \begin{equation}
 \xi(0)=\sigma_2, \xi'(0)=0, \xi''(0)=-\frac{\sigma_4}{3} 
  \end{equation}
 \begin{equation}
 \frac{\xi'(r)}{r}=-\frac{\sigma_4}{3}+...\,\,\,, \,\,\, \frac{\xi'(r)}{r}-\xi''(r) = -\frac{\sigma_6}{15}r^2 +...%{\rm for}\,  r\rightarrow 0
 \end{equation}
 so $\nabla \xi=0$ when $j=l$.
 Also as $r\rightarrow 0$, ${\cal H}\rightarrow r^2$, so  when $j=l$ in   $\nabla \nabla \xi$ the first term in the second line of Eq.\ref{eq:seconderxi} goes to zero and only terms with $p=q$ survive.
 Keeping this in mind, we can integrate  first in $\beta$. We have a quadratic form  of the type $\int d\{\beta\} \exp[-1/2 \vec{\beta}^T W \vec{\beta}+B^T\vec{\beta}]$ with a term quadratic in $\beta$ and one linear in $\beta$ that is however mixed with $\alpha$ (the expression of $B$ which depends on $\alpha$ needs some care). This means that  there is no fundamental limitation in performing analytically the integral in $\beta$, being a standard Gaussian integral, for any $N$.
 
 Clearly 
 \begin{equation}
 \sum_p \beta_{lp}\nabla_{lp}\xi_{jl}=\vec{\beta_l}\cdot ({\bf r}_j-{\bf r}_l)\frac{ \xi'(r_{jl})}{r_{jl}}
\end{equation}
\begin{equation}
 \sum_p \beta_{jp}\nabla_{jp}\xi_{jl}=-\vec{\beta_j}\cdot ({\bf r}_j-{\bf r}_l)\frac{ \xi'(r_{jl})}{r_{jl}}
\end{equation}
being zero if $j=l$.  Then 
 \begin{eqnarray}
\sum_{pq}\beta_{jp} \beta_{lq} \nabla_{jp} \nabla_{lq} \xi(r_{ij})\!\!\!&\!=\!&\!{\cal H}(r_{jl})(\vec{\beta_j}\cdot \vec{r}_{jl})(\vec{\beta_l}\cdot \vec{r}_{jl})-\vec{\beta}_{j}\cdot\vec{\beta}_{l}\frac{\xi'(r_{jl})}{r_{jl}}
%\nonumber \\
%&\!\!\!\!\!=\!&\!
=-{\cal H}(r_{jl})(\vec{\beta_j}\cdot \vec{r}_{jl})(\vec{\beta_l}\cdot \vec{r}_{lj})-\vec{\beta_j}\cdot\vec{\beta_l}\frac{\xi'(r_{jl})}{r_{jl}}
 \end{eqnarray}
 where   $\vec{r}_{jl}=({\bf x}_j-{\bf x}_l)$ and  $\vec{r}_{jl}=-\vec{r}_{lj}$. The expression for the linear term becomes:
\begin{equation}
-2 \sum_{j,l} \alpha_j\vec{\beta_l}\cdot\vec{r}_{jl} \frac{\xi'(r_{jl})}{r_{jl}} 
 \end{equation}
where   $\vec{r}_{jl}=({\bf x}_j-{\bf x}_l)$.

 In summary:
  \begin{eqnarray}
 (J, K^{-1},J)&=&\sum_{jl}\alpha_j \xi_{jl}\alpha_l %\nonumber \\
%&+& 
+\sum_{jl} {\cal H}(r_{jl})(\vec{\beta_j}\cdot \vec{r}_{jl})(\vec{\beta_l}\cdot \vec{r}_{jl})(1-\delta^K_{jl})-\vec{\beta}_{j}\cdot\vec{\beta}_{l}\frac{\xi'(r_{jl})}{r_{jl}}% \nonumber \\
%&-&
-2 \sum_{j,l}(1-\delta^K_{jl})  \alpha_j\vec{\beta_l}\cdot\vec{r}_{jl} \frac{\xi'(r_{jl})}{r_{jl}} \,.
  \end{eqnarray}
Note that $\vec {\beta}\cdot  \vec{\beta}=\beta^T \beta$. 

For simplicity let us look at the two point function.
In that case we can always choose the axis so that $\vec{r}$ is aligned with the x axis ,  so $r_{ij}=|r_{ij,x}|$, then $\vec{\beta}_j\cdot  \vec{r}_{ij} = \beta_{jx} r_{ij,x}=\pm \beta_{jx} r_{ij}$ depending on the sign of $r_{ij,x}$ .
 For the two point function we get:
 \begin{equation}
 (J, K^{-1},J)=\alpha^T \xi \alpha + \beta_x^T H'\beta_x - \beta_y^T X'\beta_y-\beta_z^T X'\beta_z-2\alpha^T Q \beta_x
 \end{equation}
 
where $H'$ elements $jl$ are $ {\cal H}(r_{jl})r_{jl}^2 (1-\delta^K_{jl})-\xi'(r_{jl})/r_{jl}$;  $X'$ elements $jl$ are $\xi'(r_{jl})/r_{jl}$ and $Q$ elements $jl$ are $-(1-\delta^k_{jl}) r_{jlx}\xi'/r_{jl}$. So 
 \begin{equation}
 P'({\bf r}_1,\ldots ,{\bf r}_N,\vec{m}) = \int d\{\alpha_j\} d\{\beta_{jp}\} \exp\left[-i\sum_{j=1}^N\alpha_j m_j-\frac{1}{2}\alpha^T \xi \alpha -\frac{1}{2}\beta_x^T H'\beta_x + \alpha^T Q \beta_x  - \beta_y^T X'\beta_y-\beta_z^T X'\beta_z\right].
 \end{equation}

 The integral in $\beta_z$ and $\beta_y$ gives $(2\pi/\det X')$. The integral in $\beta_x$ gives $(2\pi/\det H')^{1/2} \exp[1/2 \alpha^TQ H'^{-1}  Q^T\alpha]$. So for the two point function we obtain:
 \begin{equation}
 P'({\bf r}_1,{\bf r}_2,m_1,m_2)=  \frac{2\pi}{\det X'} \frac{(2\pi)^{1/2}}{(\det H')^{1/2}}\int d\{\alpha_j\}
 \exp[-i\vec{m}^T\alpha]   \exp \left[-\frac{1}{2}\alpha^T (\xi -Q^{T}H'^{-1}Q) \alpha\right]
 \end{equation}
 
 which can again be integrated giving:
 \begin{equation}
 P'({\bf r}_1,{\bf r}_2,m_1,m_2)=  \frac{2\pi}{\det X'} \frac{(2\pi)^{1/2}}{(\det H')^{1/2}}\frac{2\pi}{\det(\xi-Q^{T} H'^{-1}Q)}\exp\left[-\frac{1}{2}m^T (\xi-Q^{T}H'^{-1}Q)^{-1} m \right]\,.
 \label{eq:corrana}
 \end{equation}
 
 We can write down the matrices explicitly: 
  \begin{equation}
 \begin{matrix}
 \xi
 \end{matrix}
 =
\begin{pmatrix}
 \sigma_2  & \xi(r_{12}) \\
 \xi(r_{12}) & \sigma_2
 \end{pmatrix}
\end{equation}
 
  \begin{equation}
 \begin{matrix}
 H'
 \end{matrix}
 =
\begin{pmatrix}
 \frac{\sigma_4}{3}  & H_{12} \\
 H_{12} & \frac{\sigma_4}{3}
 \end{pmatrix}
 \end{equation}
 where 
 \begin{equation}
 H_{12}= -\xi''(r_{12})
 \end{equation}
 \begin{equation}
 \begin{matrix}
 Q
 \end{matrix}
 =
\begin{pmatrix}
 0  & -\xi'(r_{12}) \\
 \xi'(r_{12}) & 0
 \end{pmatrix}
\end{equation}

\begin{equation}
 \begin{matrix}
 Q H'^{-1} Q
 \end{matrix}
 = \frac{1}{(\frac{\sigma_4}{3})^2-H_{12}^2}
\begin{pmatrix}
  -\frac{\sigma_4}{3}\xi'^2 (r_{12})& -H_{12}\xi'^2 (r_{12}) \\
 -H_{12}\xi'^2 (r_{12}) & -\frac{\sigma_4}{3}\xi'^2 (r_{12})
 \end{pmatrix}\,.
\end{equation}
If we define $G$ as $\xi-QH'^{-1}Q$ we have that:
\begin{equation}
\begin{matrix}G^{-1}
\end{matrix}
=
\begin{matrix}
( \xi-Q H'^{-1} Q)^{-1}
 \end{matrix}
=\left[\xi^{-1}\det[\xi]+\frac{\xi'^2(r_{12})}{\left(\frac{\sigma_4}{3}\right)^2-H_{12}^2}\right.
\begin{pmatrix}
\frac{\sigma_4}{3} &-H_{12}\\
-H_{12}&\frac{\sigma_4}{3}
 \end{pmatrix}
\bigg]{\cal N} 
\end{equation}
where 
\begin{equation}
{\cal N}=\left(\sigma_2+\frac{H_{12}\xi'^2(r_{12})}{\left(\frac{\sigma_4}{3}\right)^2-H_{12}^2}\right)^2-\left(\xi_{12}+\frac{\frac{\sigma_4}{3}\xi'^2(r_{12})}{\left(\frac{\sigma_4}{3}\right)^2-H_{12}^2}\right)^2\,.
\end{equation}
Then
\begin{equation}
\det G = \frac{1}{\left(\frac{\sigma_4}{3}\right)^2-H_{12}^2} \left[  \left(\frac{\sigma_4}{3}\right)^2 \left(\sigma_2^2-\xi^2_{12}\right)+H_{12}^2(\xi_{12}^2-\sigma_2^2)+2\frac{\sigma_4}{3}\sigma_2\xi'^{2}_{12}-2H_{12}\xi_{12}\xi'_{12}+\xi'^{4}_{12}\right] \nonumber
\end{equation}
where we have used the shortcut $\xi_{12}=\xi(r_{12})$. This leaves only the $m$ integral to be performed which gives something akin to an error function. However, note that for large thresholds above $\sim 3\sigma$ it will not be needed to integrate as the integral is dominated by the value of the threshold ($m$). 

Eq.~\ref{eq:corrana} is our main result and provides an exact result for the correlation function of extrema of a random gaussian field, weighted by $1/|\det w|$.
As anticipated above, to compare this expression to observations, the data must be suitably weighted. Moreover for this expression to be a good description of the peaks clustering properties the  threshold must be sufficiently high. For Gaussian fields  this applies for thresholds above $2\sigma$, which is most relevant for cosmological applications.

\begin{figure}
\centering
\includegraphics[width=0.7\columnwidth]{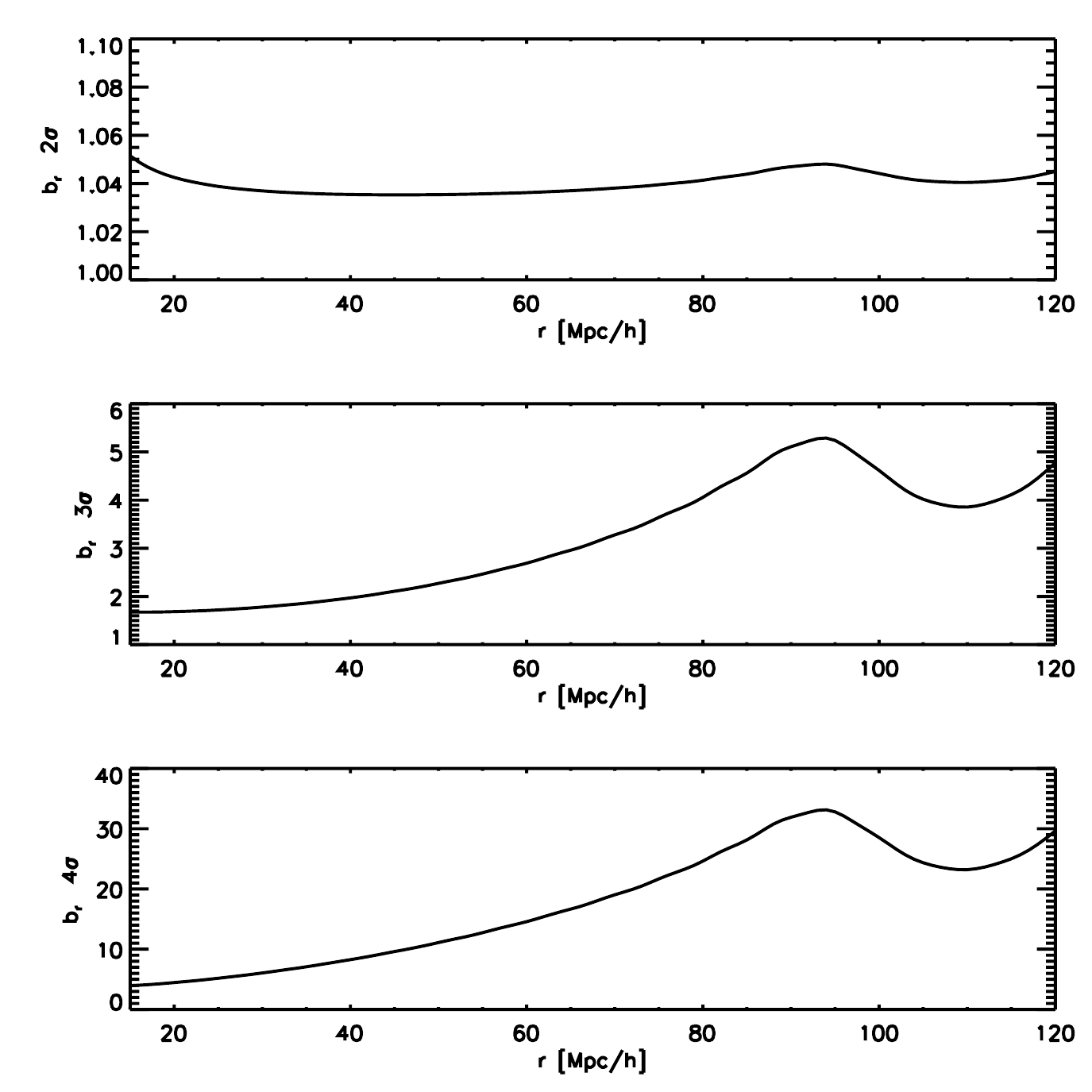}
\caption{The bias  $b_r$ defined as  the square root of the ratio of the extrema correlation function to the dark matter one according to Eq.~\ref{eq:corrana}. Here we show  (from top to bottom panel) 2,3, and 4 $\sigma$ extrema. Note the  increased boost as function of threshold, the broad-band scale-dependent bias and the localised ``bump" in the bias at around $r=90$ Mpc/h.}
\label{fig:eq28}
\end{figure}

Figure \ref{fig:eq28}   shows the bias  $b_r$ defined as  the square root of the ratio of the extrema correlation function to the dark matter one (where we have used the millennium simulation dark matter correlation function as input, see next section) according to Eq.~\ref{eq:corrana}:
\begin{equation}
b_r\equiv \sqrt{\frac{\xi_{\rm ex}(r)}{\xi_{\rm DM}} }\,.
\end{equation}
 This figure shows several effects:  {\it a)} an overall correlation boost that increases with increasing threshold as expected,  {\it b)} a  broad-band scale-dependent bias which becomes more marked with increasing threshold, also as expected e.g.  in the peak background split , {\it c)} a new feature of a relatively narrow, scale-dependent feature in the bias (a ``bump") which is located very near the Baryon Acoustic Oscillation feature. 
 The broadband effect is likely to be  affected by the evolution bias (see \S \ref{sec:evbias}) or by the choice of weighting scheme, but   the localized  feature is very robust to these effects. This feature is particularly interesting and has been mostly  overlooked in the literature, but see \citet{Desjacques:2008}.  Mathematically,  the BBKS  derivation of the correlation function of maxima assumes that derivatives of the  two point correlation function of the density field can be ignored. On the other hand  the presence of the BAO  signal introduces a changing first and second derivative of the correlation function which  are responsible for non-negligible effects. In other words, selecting peaks of a Gaussian field is a highly non-linear operation which creates a highly non-Gaussian  peaks field. The essence of non-Gaussianity  is mode coupling which, by moving power across scales, tends to move  around, distort  or even erase localized features.   Below we will test the performance of Eq.~\ref{eq:corrana} both on Gaussian random fields and on simulations and  examine the broadband and the localised effects separately. 
 
\section{Validation and tests on Numerical simulations}
\label{sec:validation}
We use two sets of simulations: 
\begin{itemize}
\item{} 20  random realisations of a Gaussian field cast in a $1 h^{-1} \mathrm{Gpc}$ box, divided into $512^3$ cells,  with power spectra matching that of linear theory for the same cosmological parameters as used in the Millennium-I simulation. These are used to test the accuracy and validity of our analytic expression and  weighting   choices.  
\item{} The dark matter outputs from the Millennium-I simulation \citep{Millenium1:2005,gadget2,Millenium2:2006}.
The 
simulated box size is $500 h^{-1} \mathrm{Mpc}$ comoving and the  adopted cosmology  is   not too dissimilar   from the concordance $\Lambda$CDM:  a spatially flat universe with matter density parameter $\Omega_m=0.25$,  baryon density parameter $\Omega_b=0.045$, scale invariant primordial power spectrum, Hubble constant $H_0=73$km/s/Mpc and {\it rms} of fluctuations of 8 $h^{-1}$ Mpc scale $\sigma_8=0.9$. Individual particles have a mass of $8.6\times 10^8$ $h^{-1}$ M$_{\odot}$.
 We  employ three snapshots  at redshifts of $z = 0, 0.687, 127$  the snapshot at $z=127$ being the initial one. The density fields, in real space, are constructed by assigning the $2160^3$ particles to a $256^3$ grid.  \end{itemize}
 
 The $z = 127$ snapshot from the   Millennium-I simulation  is well described as a Gaussian Random Field. Having several realizations of Gaussian fields reduces the variance and therefore errors  especially on large scales. On the other hand, when studying the evolution bias,  taking ratios between different snapshots of the same simulation reduces cosmic variance errors.
 
 In both cases a grid point is identified as extreme if along each of the three spatial axes the two adjacent grid points both lie at higher or lower densities than the value of the grid point. Local maxima are the special case of extrema where the grid point is greater than all six adjacent points. Only extrema with densities above particular thresholds are considered, and for higher thresholds the extrema are predominantly maxima.
This procedure effectively defines the $n-\sigma$ threshold for  smoothing scale  corresponding to the cell size of $1.9$ Mpc/$h$.
The power spectra and correlation functions measured from the simulation and the realisations  for a $4\sigma$ threshold is very noisy and  sometimes not informative, therefore it will not  always be displayed.

 \subsection{Validation on  Gaussian fields}
\label{sec:valid:gaussian}

\begin{figure}
\centering
\vspace*{-0.0cm}
\includegraphics[width=0.475\columnwidth]{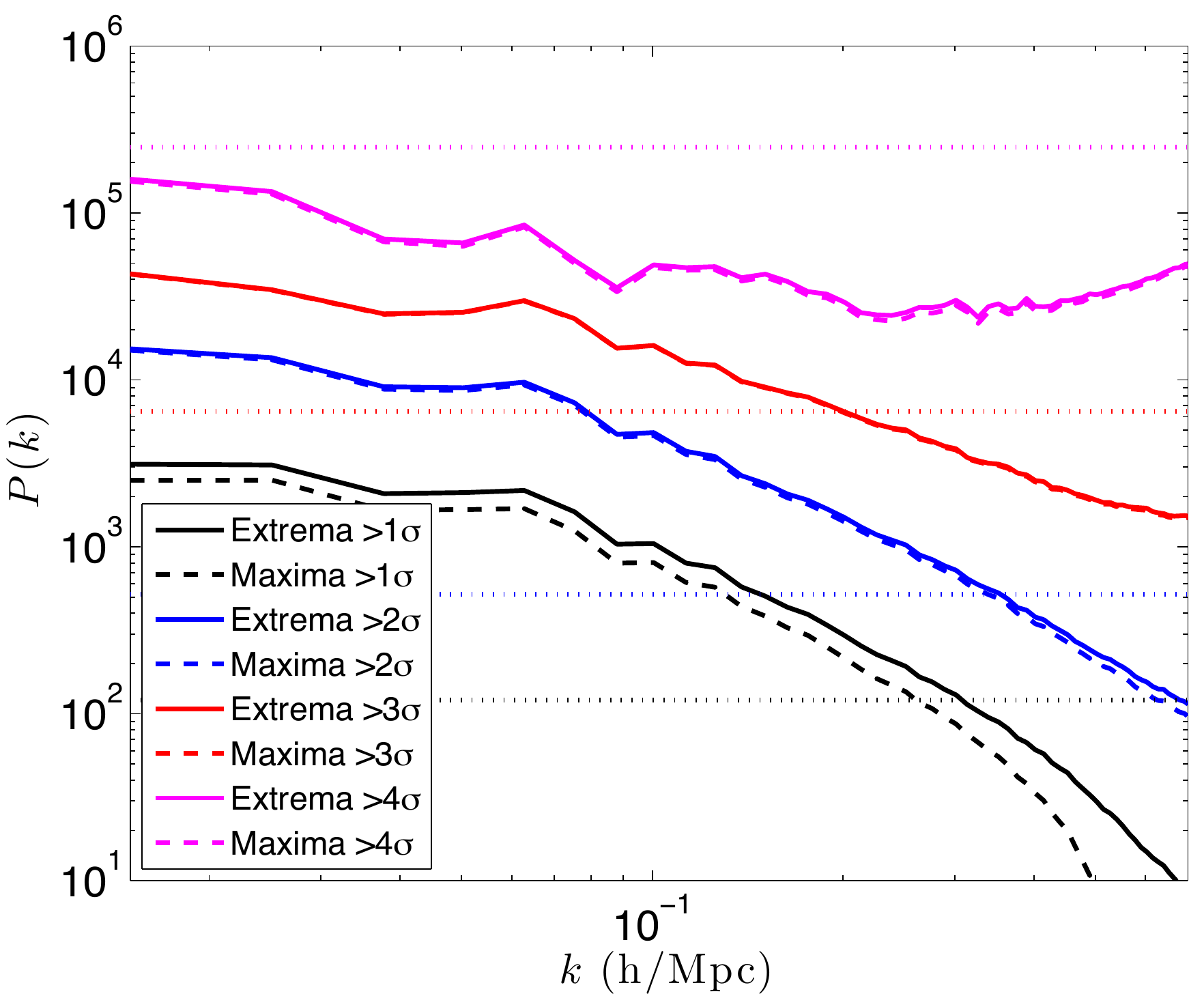}
\includegraphics[width=0.45\columnwidth]{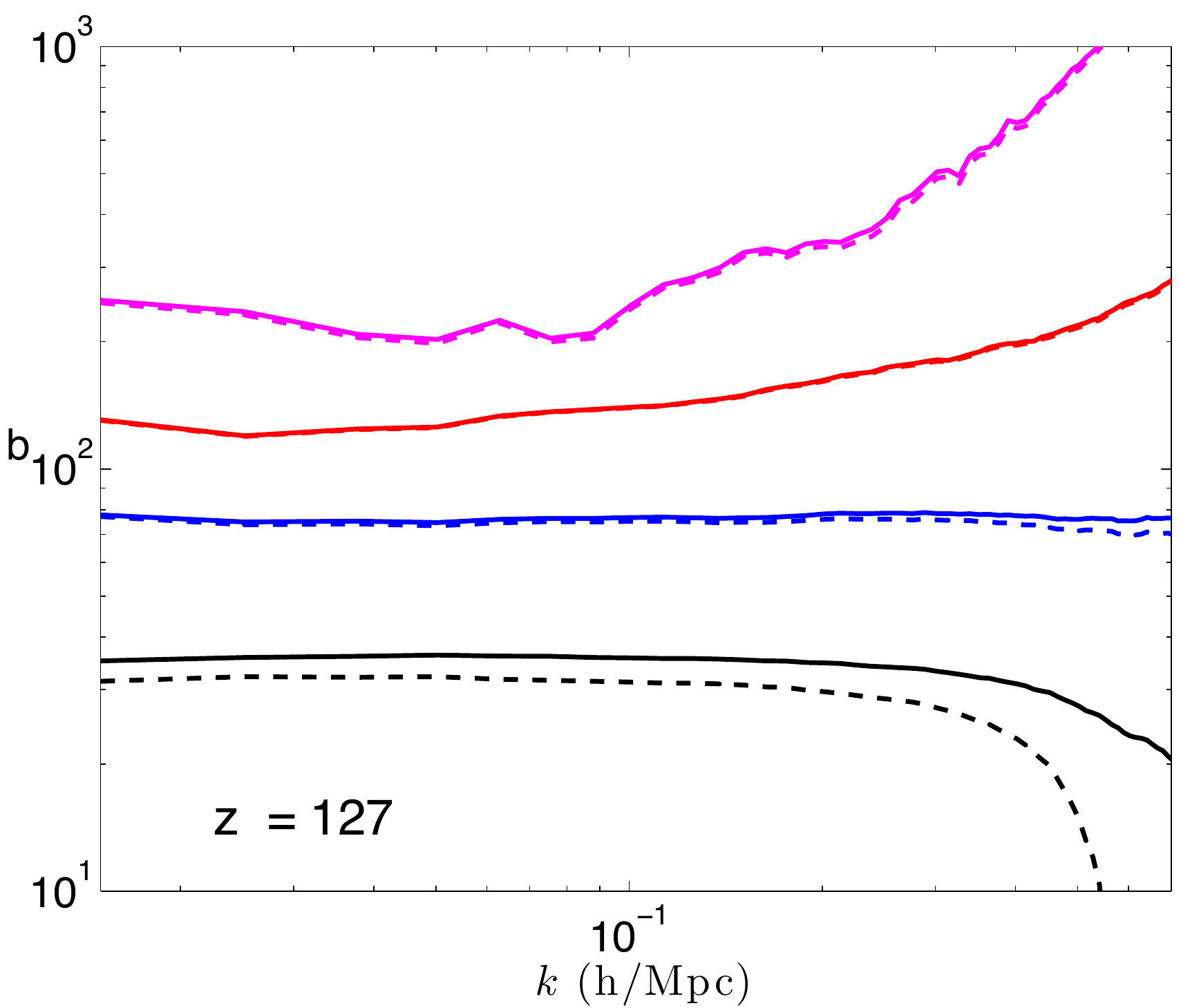}
\caption{Left: power spectra of extrema and maxima in the $z=127$ snapshot from the Millennium I simulation in real space. Poissonian shot noise has been subtracted and its magnitude is illustrated in each case by horizontal dotted lines. Top-Right: the bias $b_k$ obtained from the ratio of the ``peaks" (extrema and maxima) to matter power spectra.} 
\label{fig:maxima}
\end{figure}

We begin by analyzing the $z = 127$ snapshot.  In Fig.~\ref{fig:maxima} (top left panel)  the power spectra of the extrema and maxima are shown as solid and dashed lines respectively. Each line has been corrected for the window function associated with the grid, and for Poissonian shot noise.  The same power spectra are recast in the form of the bias parameter in the top right panel of Fig.~\ref{fig:maxima}. Here and in what follows the bias $b_k$ is defined as
\begin{equation}
b_k \equiv \sqrt{\frac{P_{\rm ex,pk} (k,z)}{D^2(z) P_{\rm DM} (k,z =127)}}
\end{equation}
where $D(z)$ denotes the linear growth function and the subscripts exp, pk denote extrema and peaks respectively.

For completeness in Table \ref{tab:numbers} we report the number of peaks and extrema located in the three snapshots of the Millennium I simulation as a function of the threshold. For  thresholds above $2\sigma$, extrema are to a good approximation maxima and the error that the extrema/maxima approximation introduces on the bias is below $5\%$ on large scales ($k < 0.4 \hmpc$). This demonstrates that the extrema approximation for maxima works extremely well.

\begin{figure}
\centering
\vspace*{-0.0cm}
\includegraphics[width=0.454\columnwidth]{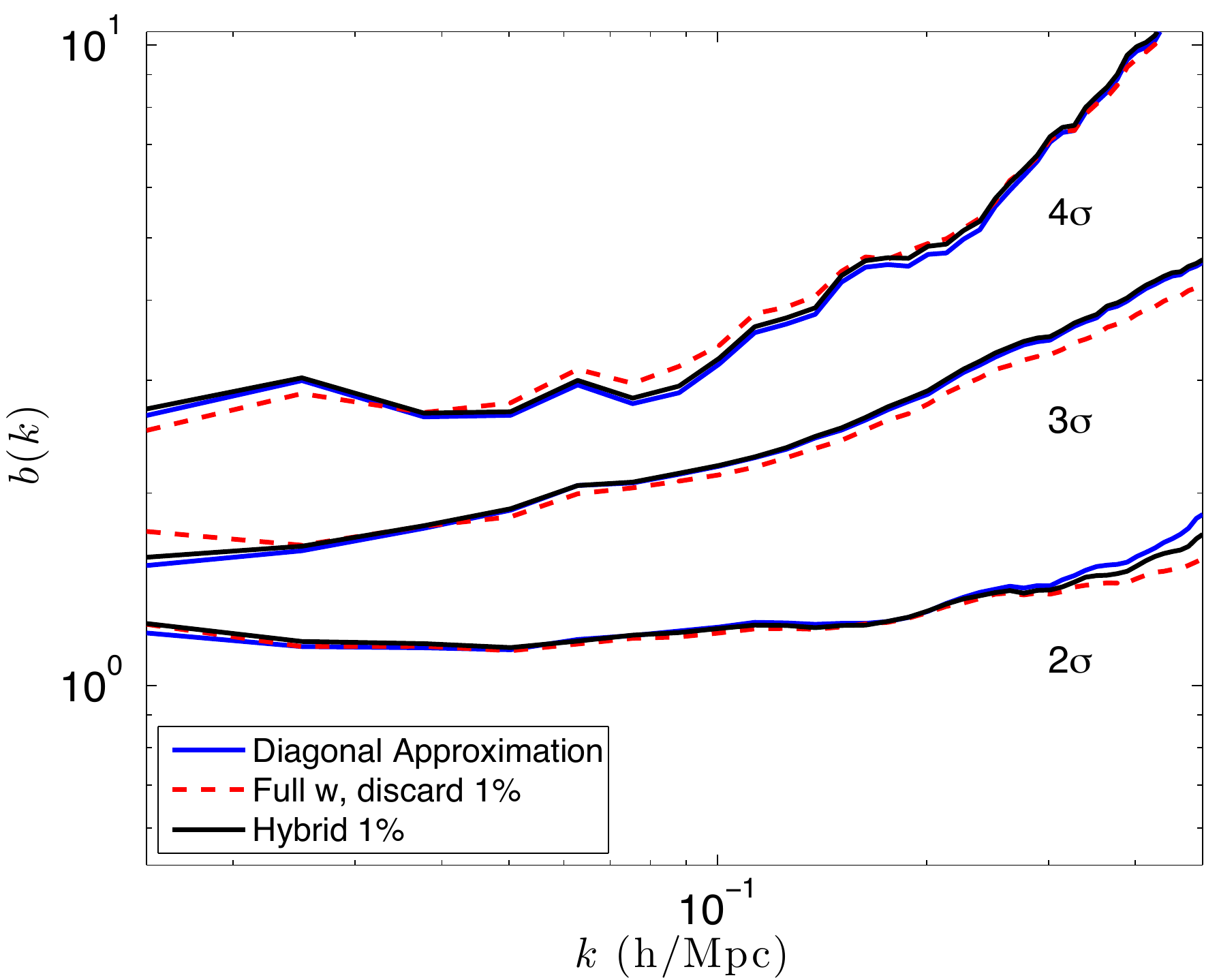}
\includegraphics[width=0.475\columnwidth]{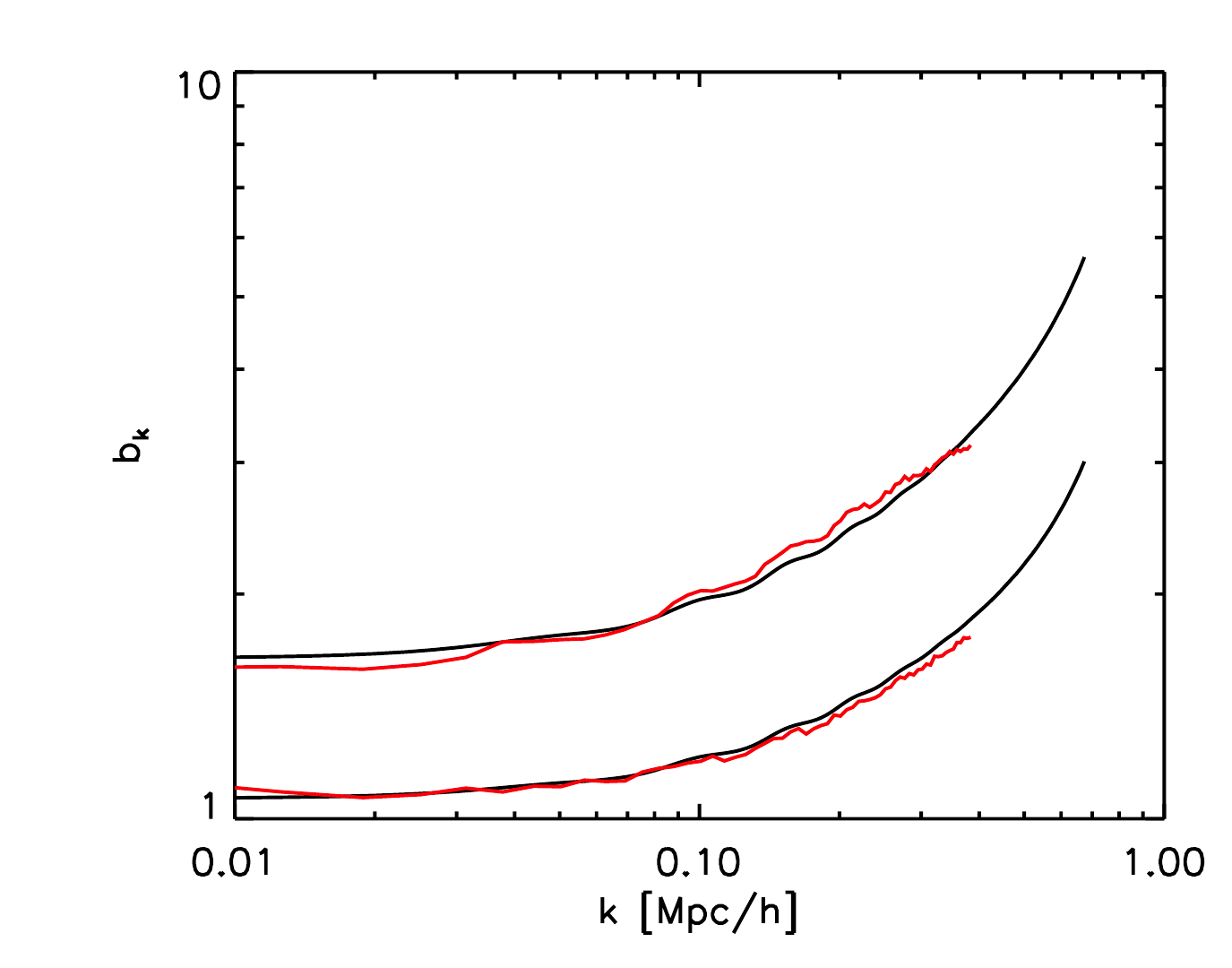}
\caption{Left:  bias $b_k$ obtained from the average of the 20 Gaussian realizations for different approximations to the $|\det w|$ weighing, see text for more details. Right:  $b_k$ for $2$ and $3 \sigma$ thresholds, from the theoretical prediction obtained by Fourier-transforming  Eq.~\ref{eq:corrana} and from the Gaussian realisations.}
\label{fig:weidetw}
\end{figure}

In the analytical formulation, by   omitting the term given by $| \det w_{(j)}|$ in (\ref{eq:path})  we arrive at the correlation function associated with peaks which have been assigned weights given by  $1/| \det w_{(j)}|$. 
Before we can compare simulation outputs with the analytic results we need to find a numerically stable way to compute the quantity $|\det w|$ which involves the numerical evaluation  via finite differences of six second derivatives  around each peak. As the weighting we will employ will be $1/|\det w|$, when $|\det w|$ is small, any numerical error gets amplified. We have tried several different approximations to tame this effect, in particular: considering only diagonal elements in $w$ (i.e., off-diagonal elements are taken to be zero which corresponds to the assumption of spherically symmetric peaks);
 computing the full $w$ matrix but excluding from the analysis the 1\% of peaks with the smallest $|\det w|$  and an hybrid approach where the full $w$ matrix is computed on the 99\% of the peaks with the largest $|\det w|$ and a diagonal approximation implemented in the remaining 1\%.  The bias $b_k$ obtained for these three approaches\footnote{The weighting scheme boosts broad peaks (small $|\det w|$) so neglecting even only 1\% of the broad peaks can have a non-negligible effect, that is  why the approximation that discards completely these peaks  does not work too well.} is shown in the left panel of Fig. \ref{fig:weidetw}. The diagonal approximation works very well and is what we will use  in what follows. 

In the right panel of Fig. \ref{fig:weidetw} we show the average for $b_k$ for peaks above 2 and 3 $\sigma$,  of  the 20 Gaussian realizations and the theoretical prediction which has been obtained by Fourier-transforming Eq.~\ref{eq:corrana}.

From the right panel of Fig. \ref{fig:weidetw}  we can identify  the same effects mentioned above:
{\it a)} a broad band  scale-dependent bias which  implies a broadband change of shape  of the peaks/extrema power spectrum compared to the underlying/dark matter one and  {\it b)} the localized  effect in real space (the ``bump" in $b_r$ at $r\sim90$ Mpc$/h$); this is a periodic feature in $b_k$ which will be examined in  Sec.~\ref{sec:BAO}.

\subsection{Tests on low z snapshots of N-body simulations}
The lower redshift snapshots correspond to significantly non-Gaussian fields.
In order to identify equivalent threshold levels to those found in the $z=127$ snapshot we apply the condition that the volume fraction of the field lying above the threshold matches that of the Gaussian case. In practice this is achieved by applying a Gaussianisation transformation to the field \citep{1992Weinberg}. This involves applying a weighting such that the one-point distribution of the resulting field is Gaussian, while the rank order of the field is preserved. This allows the sigma thresholds to retain their original meaning, and ensures the threshold encompasses the same volume fraction, but may produce different results compared to defining a threshold by selecting a fixed number of peaks. 
In what follows we will always refer to this procedure when we consider extrema above a threshold.
The numbers of extrema above several thresholds are reported in table~\ref{tab:numbers}.

\begin{figure}
\centering
\vspace*{-0.0cm}
\includegraphics[width=0.45\columnwidth]{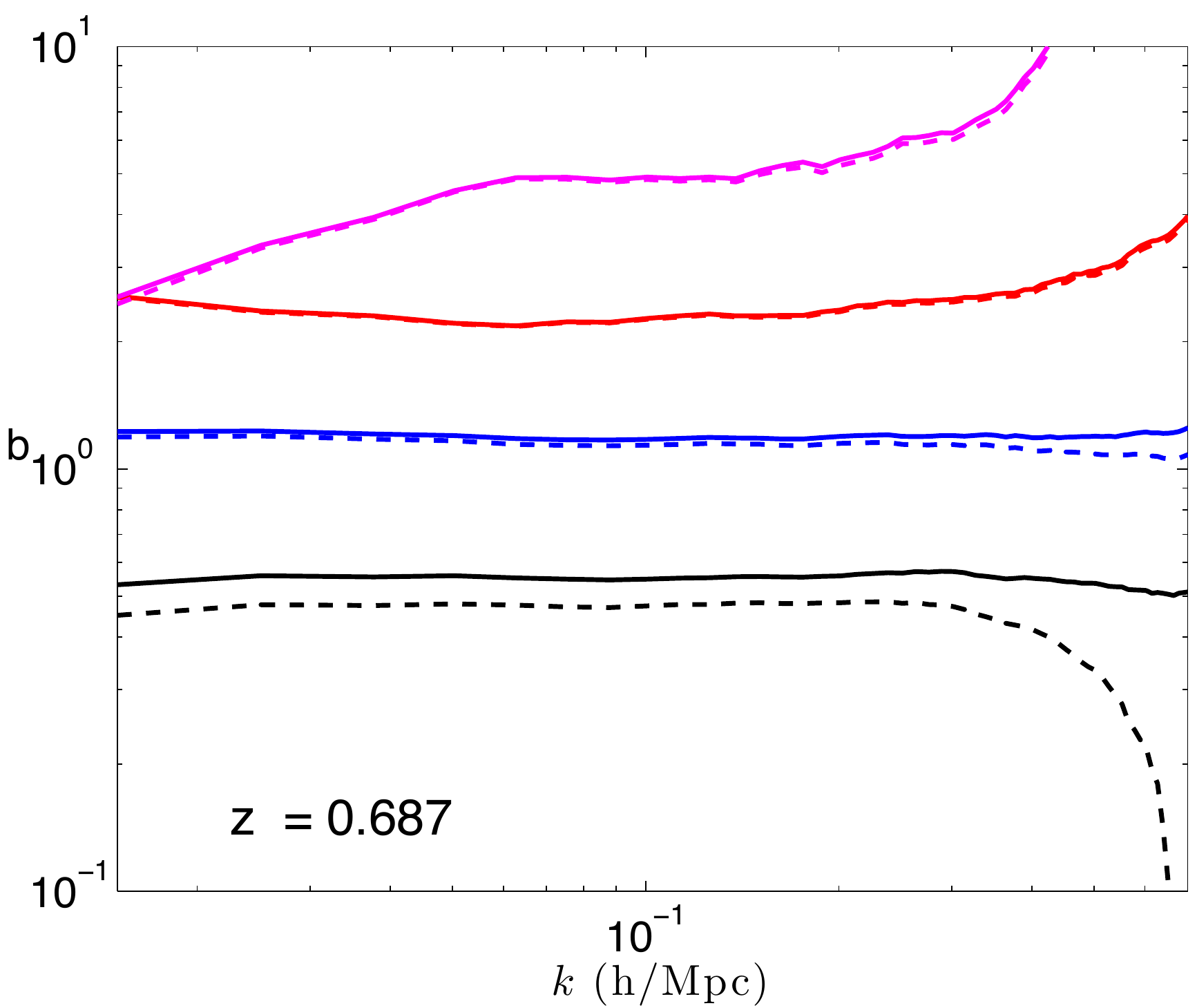}
\includegraphics[width=0.45\columnwidth]{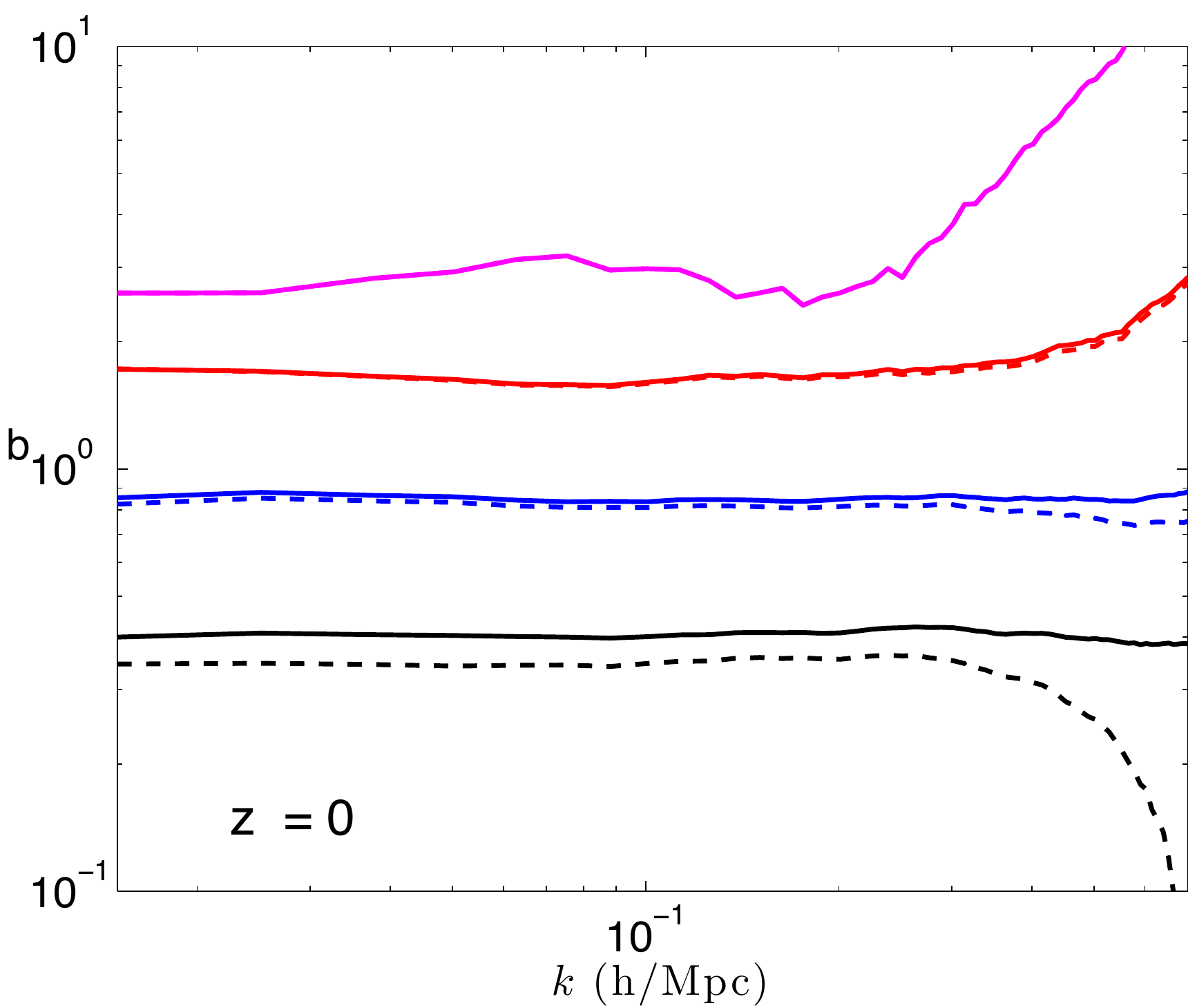}
\caption{The bias $b_k$ obtained from the ratio of the ``peaks" to matter power spectra. Left: extrema and maxima from the $z=0.687$ snapshot. Right: same  but using the $z=0$ snapshot.}
\label{fig:maximalowz}
\end{figure}

Fig. \ref{fig:maximalowz} shows the bias $b_k$ for the $z=0.68$ and $z=0$ snapshots. As before we see that  for thresholds above $2\sigma$ the identification of extrema with maxima is a very good approximation.

 \begin{table} 
 \caption{Number of peaks and extrema as function of the threshold }
 \label{tab:numbers}
 \begin{center}
 \begin{tabular}{@{}lcccccc}
  \hline
  snapshot's z & threshold & mass&extrema& maxima& \% bias difference\\
 $127$        &  $1-\sigma$& & 985734  & 919000 &\\
  $$       &  $2-\sigma$ && 260454       & 256463 & $<5\%$ for $k<0.4$\\
  $$       &  $3-\sigma$ && 27068         & 27011 & $<1.5\%$\\
  $$       &  $4-\sigma$ && 952            & 951 & $<1\%$\\
  \hline
  % extrema: 884895 209216 17265 487
%  maxima: 777869 202339 17205 486
  $0.68$       &  $1-\sigma$ && 884895 & 777869 &\\
  $$       &  $2-\sigma$ &      & 209216 & 202339 & $<10\%$ for $k<0.4$\\
  $$       &  $3-\sigma$ &    & 17265 & 17205 & $<2\%$\\
  $$       &  $4-\sigma$ &    & 487 & 486 & $<5\%$ \\
   \hline
%  extrema: 854610 198457 16350 472
% maxima: 748939 191877 16279 472
  $0$       &  $1-\sigma$ && 854610 & 748939 &\\
  $$       &  $2-\sigma$ && 198457 & 191877 &  $<10\%$ for $k<0.4$\\
  $$       &  $3-\sigma$ && 16350 & 16279 & $<5\%$ \\
  $$       &  $4-\sigma$ && 472 & 472 &  $<1\%$\\
  \hline
  \end{tabular}
   \end{center}
  \end{table}
 
\begin{figure}
\centering
\includegraphics[width=0.45\columnwidth]{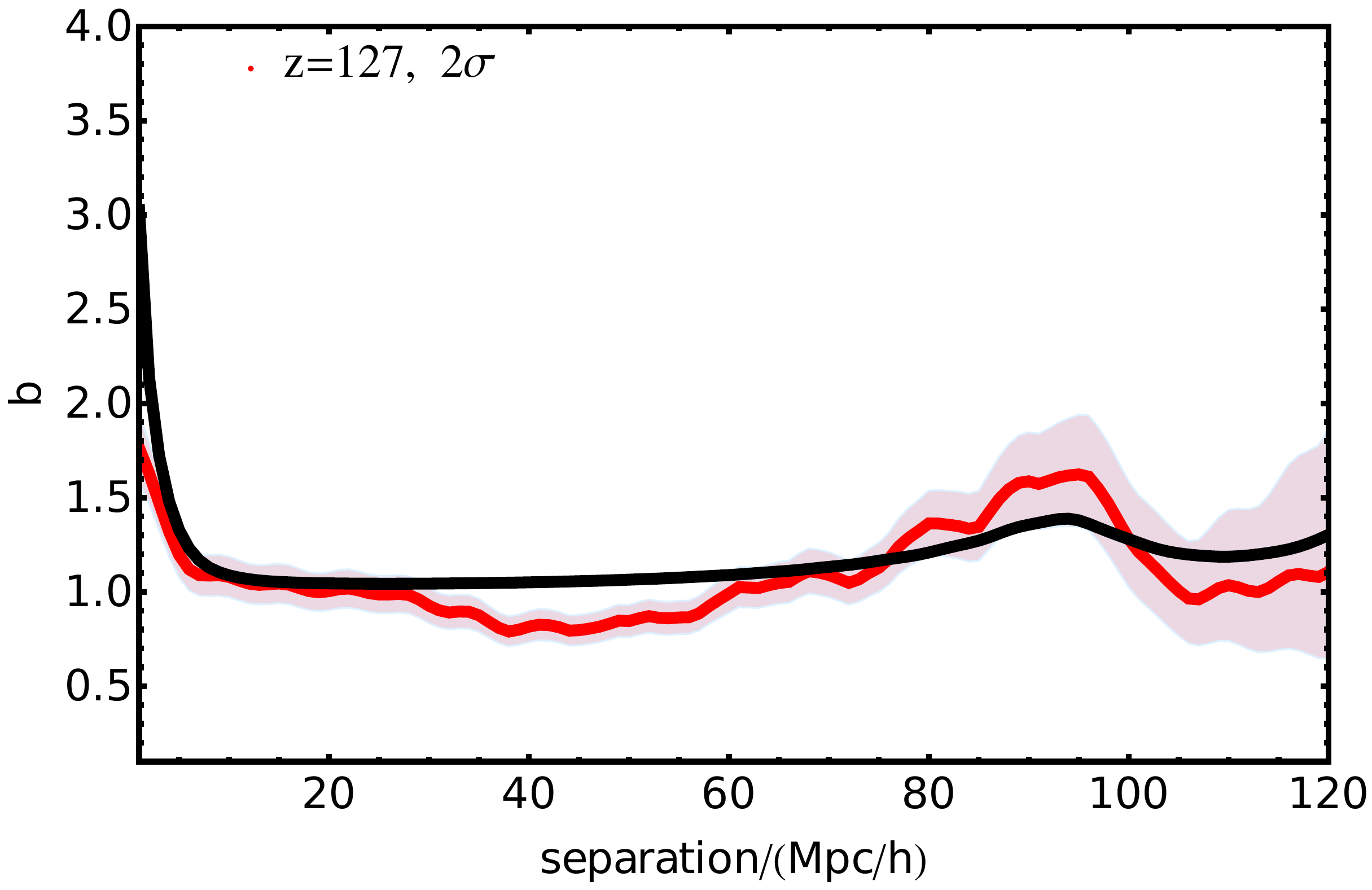}
\includegraphics[width=0.45\columnwidth]{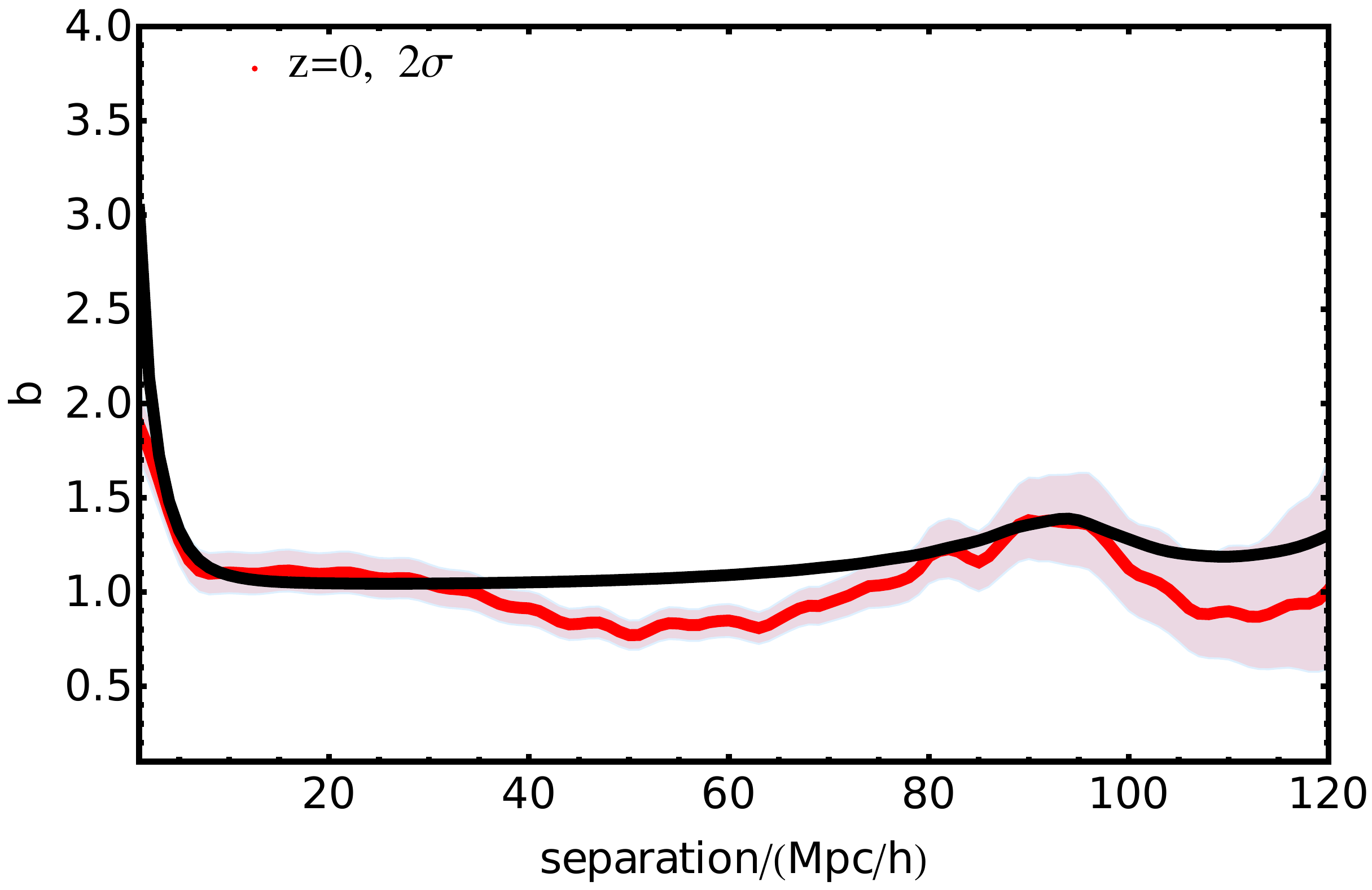}
\includegraphics[width=0.45\columnwidth]{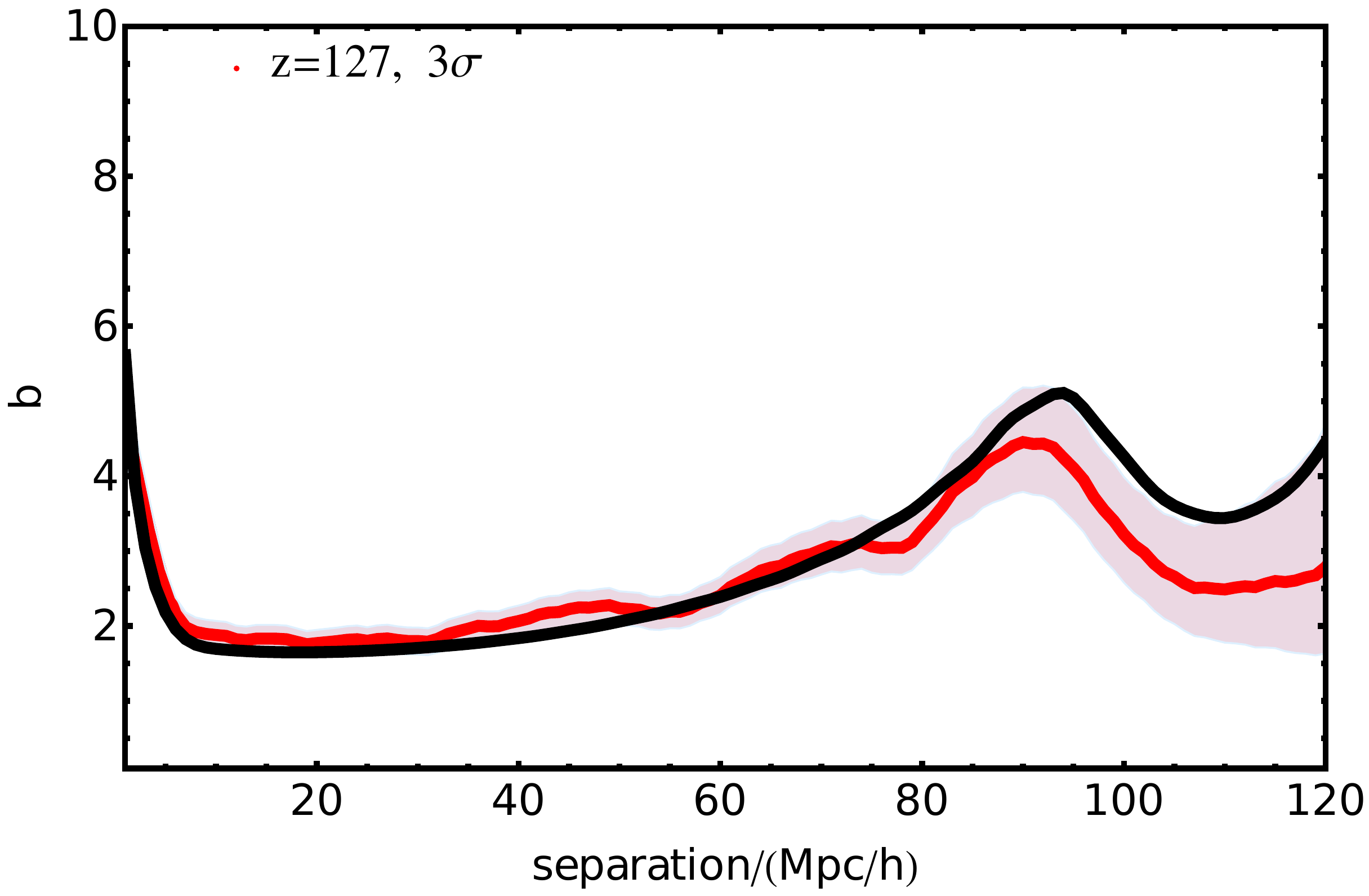}
\includegraphics[width=0.45\columnwidth]{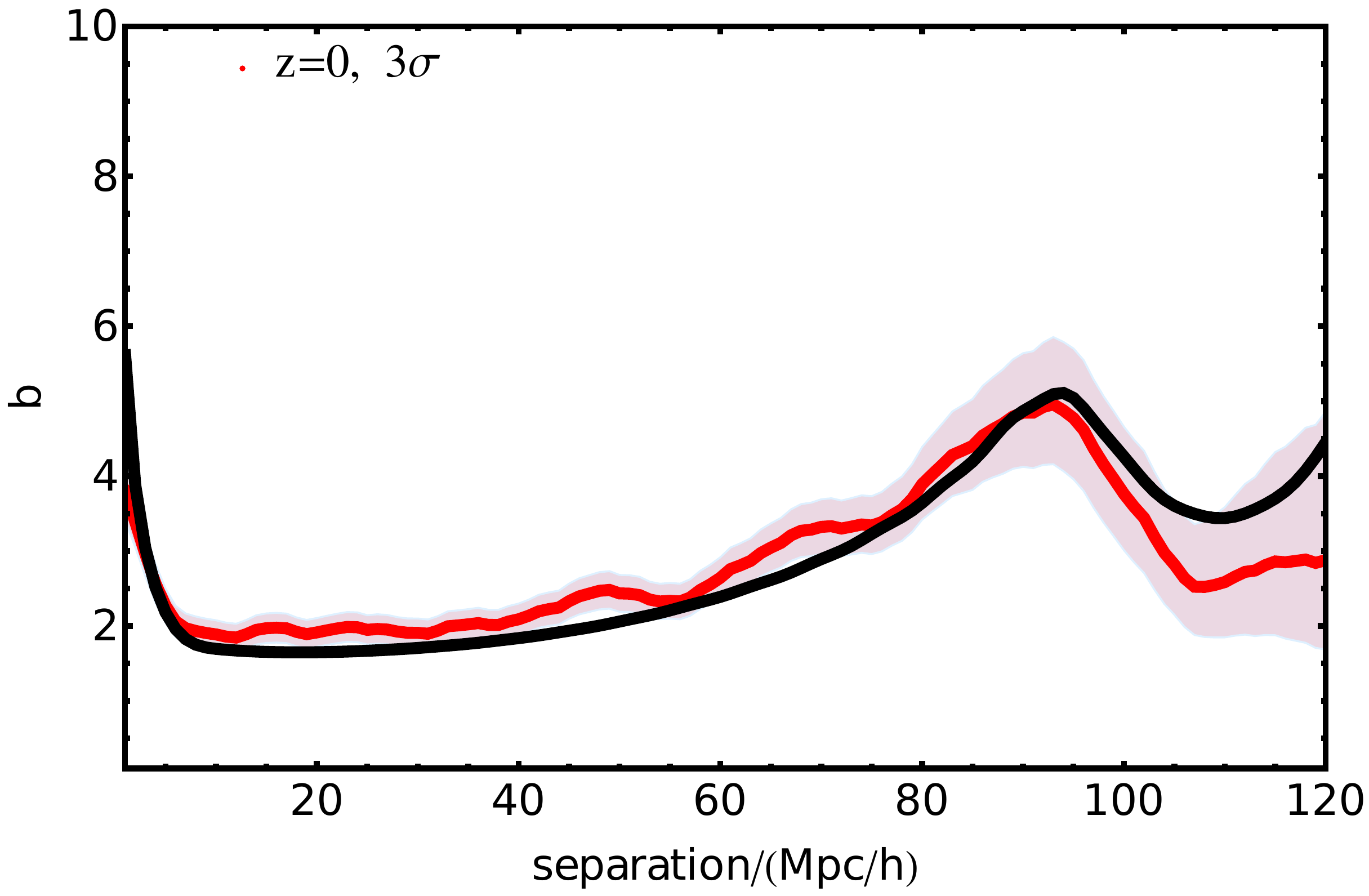}
\caption{The bias of peaks as a function of scale, top panels for the  $2 \sigma$ threshold  and bottom panels for $3 \sigma$ threshold.
Left column is at $z=127$, right column at $z=0$. Black line is the theory prediction from Eq.\ref{eq:corrana},  red line  corresponds to the peaks of the  Millennium simulation. The weighting used is $1/|\det w|$.}
\label{fig:bias}
\end{figure}

Figure  \ref{fig:bias} shows the comparison between the theory prediction of Eq.~\ref{eq:corrana} for $b_r$ and  for the peaks of the Millennium simulation.  Top (bottom) panels are for $2 \sigma$ ($3 \sigma$) threshold; left (right) panels are for $z=127$ ($z=0$).
The weighting used is $1/|\det w|$. The theoretical prediction reproduces well the low redshift snapshots   for scales $<100$ Mpc/$h$, with a maximum deviation  of $<20\%$   for $3 \sigma$ peaks at $z=0$. It reproduces qualitatively the bias behaviour at larger scales. Note however that  given the size of the simulation, cosmic variance is large on those scales.   

 From this comparison (left vs right panels) we can also appreciate that the evolution bias is small.

From Tab.~\ref{tab:numbers} it is apparent that the choice of threshold based in the Gaussianisation procedure adopted does not conserve the number of peaks.  If it is true that to  a good approximation, peaks can be identified with  halos and halos do not merge, the number should be conserved.
On might therefore worry that the selection adopted introduces an error in the threshold; this is then interpreted like evolution bias.
We have quantified this by also selecting thresholds at lower redshifts so that the number of peaks is conserved (i.e. is the same as  at $z=127$) and computed the same quantities. The  difference in $b_k$ between the  constant peak numbers and the constant number of $\sigma$ of the Gaussianised field is  at the 6\% level and is constant on large scales ($k<0.6$ $h/$Mpc).

\subsection{Quantifying Evolution bias}  
\label{sec:evbias}
\begin{figure}
\centering
\includegraphics[width=0.45\columnwidth]{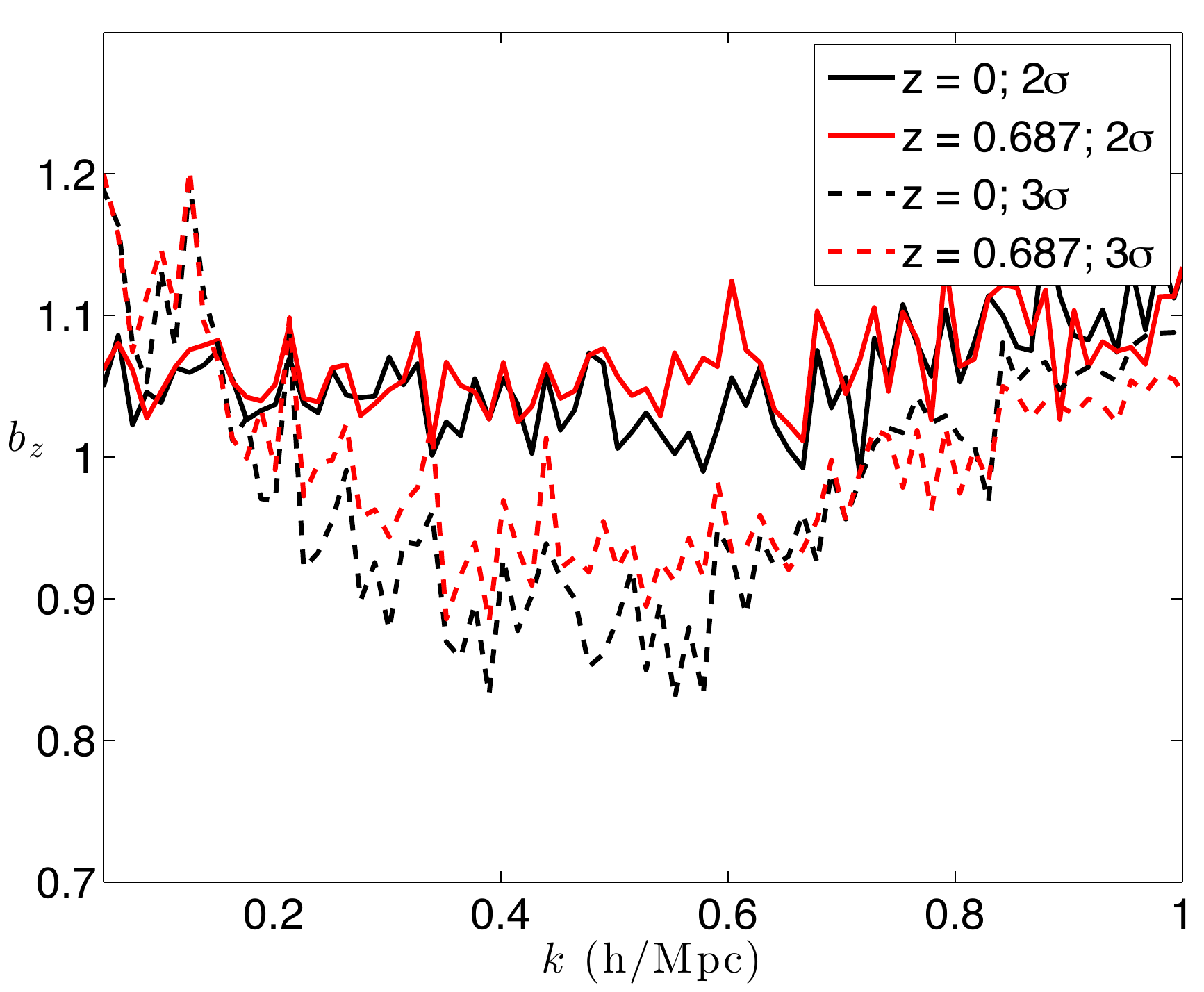}
\includegraphics[width=0.45\columnwidth]{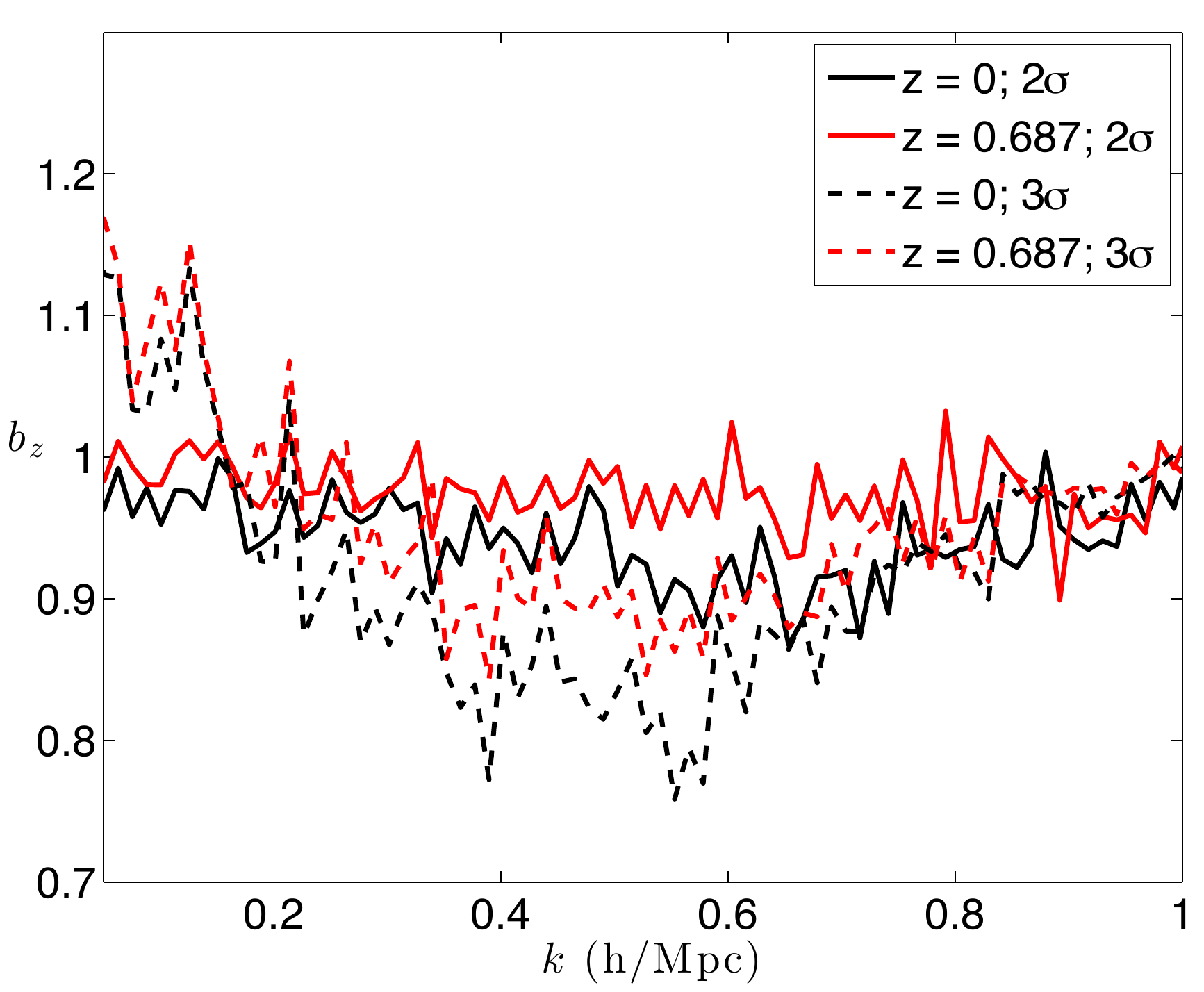}
\caption{The evolution bias, as defined in Eq.~(\ref{eq:zBias}), as a function of wavenumber. The peak power spectra $P_{\rm pk} (k,z)$ at $z=0$ (black)   and $z=0.68$  (red) show little sign of evolution with respect to the reference  $z=127$ field. The thresholds considered here to define the minimum peak height are $2 \sigma$ (solid lines)   and  $3 \sigma$ (dashed), left using the Gaussianization procedure, right keeping the number of peaks fixed at the value of $z=127$.}
\label{fig:zBias}
\end{figure}

In Fig \ref{fig:zBias} we compare the power spectra of  peaks for the two evolved snapshots, at  $z=0.687$ and $z=0$, with the power spectrum associated with the peaks of the Gaussian snapshot at $z = 127$. We define evolution bias as  $b_z$:
\begin{equation}
b_z \equiv \sqrt{\frac{P_{\rm pk} (k,z)}{P_{\rm pk} (k,z = 127)}} \, .
\label{eq:zBias}
\end{equation}
 This evolution bias  remains  relatively close to unity, illustrating how little the spectral power of the peak field has changed, despite the underlying density field growing by approximately two orders of magnitude. The maximum excursion from unity is $\sim15\%$ and is at non-linear scales ($k\sim0.6 h$/Mpc). Modelling in detail the evolution bias goes beyond the scope of this paper, other works in the literature address this  specifically e.g.,  \cite{Fry:1996,Tegmark/Peebles:1998,evbiasHui,evbiaspercival}.
 
The issue of the correspondence of peaks to halos, although one of the main pillars of peak theory and other modelling approaches and one of the main assumptions used here, is not simple and straightforward. As already mentioned, \citet{Ludlow/Porciani:2011} find that a high percentage of halos correspond to peaks of the initial density field. More recently \cite{Rubin/Loeb:2013}  argue that there is a close peak-halo relation but the definition of the threshold might affect the correspondence (in our language this could be taken care of with a mapping of the threshold in terms of $\sigma$ vs mass of the halo). Many halo finder algorithms which are applied to N-body simulations to identify halos (e.g., SO, AHF), work under the assumption that halos are density  peaks, but these are not guaranteed to correspond to the peaks of the initial density field.
We find that, in agreement with the works mentioned above, there is a peak-halo correspondence, which opens up the possibility that the conclusions we have drawn for peaks  can be applied to halos. 

The $1/|\det w|$ weighting that we applied to peaks cannot however easily be applied to halos. Moreover at $z\sim 0$  it is likely that non-linear evolution has changed the shape of the peaks so much that  $\det w (z=0)$ and  $\det w (z \gg 0)$  computed at the same position might be very different. Therefore  a proxy must be found. We  start addressing this in the appendix although this is still somewhat an open issue. 
More critically we find that the broad band shape of the  correlation  function (and power spectrum)  of halos change drastically  with different choice of weighting. For example the  power spectrum of halos weighted by halo mass at large scales  ($k<0.1 h$/Mpc) coincides with the prediction of Eq. \ref{eq:corrana}. Its  broad band shape however is much closer to that of the linear matter power spectrum at $0.1<k {\rm [{\it h}/Mpc]}<0.4$.   On the other hand a weighting of halos by  the inverse of their mass --which  is one of our initial {\it ansatzes}  for a $1/|\det w|$ proxy applicable to halos-- yields a much closer description to the shape of the peaks power spectrum at the same redshift although the amplitude is $\sim 20\%$ lower.

\begin{figure}
\centering
\includegraphics[width=0.5\columnwidth]{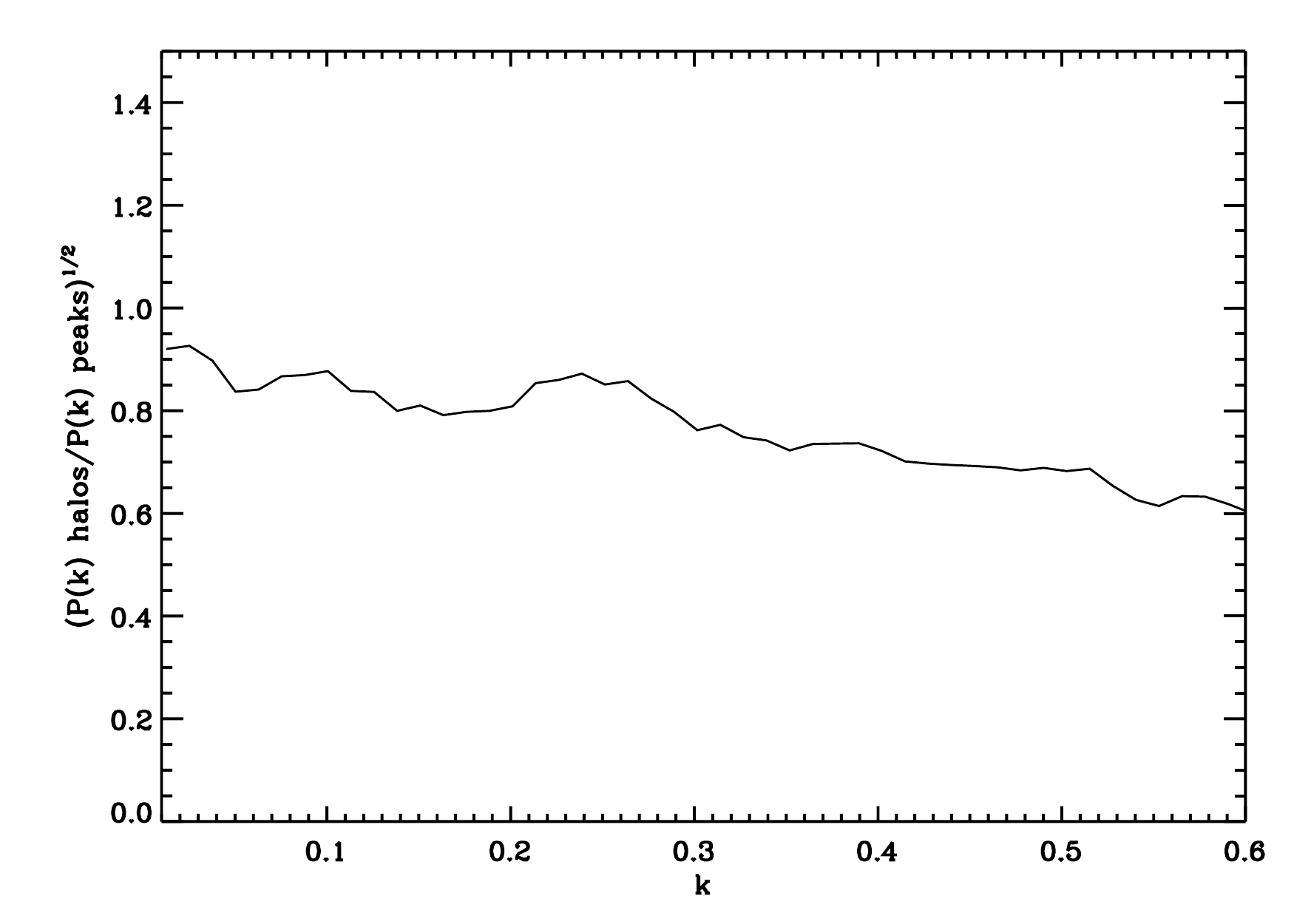}
\caption{The bias of  halos at $z=0$ (weighted by the inverse of their mass, our proxy for the  $1/|\det  w|$) with respect to peaks at the same redshift weighted by $1/|\det w|$. The threshold  is $3~\sigma$ (see text for more details).}
\label{fig:biashalos}
\end{figure}

In Fig. \ref{fig:biashalos} we show the bias of halos at $z=0$ (weighted by the inverse of their mass; which  is our initial {\it ansatze}) with respect to peaks at the same redshift weighted by $1/|\det w|$.  The halos are selected in a similar way to the peaks, by considering them only if the coincide with the region above the threshold. The selected threshold here is $3~\sigma$ which here corresponds to a minimum mass of $2.4\times 10^{11}$M$_{\odot}/h$. The relative  bias shows maximum deviations of order 20\% but is clearly a smooth linear dependence on $k$; much of this mis-match  could  be corrected for by  adjusting the weighting and/or the threshold (as motivated e.g., by \cite{Rubin/Loeb:2013}). For example, for the case of $3 \sigma$ halos at $z=0$ it is sufficient to use $2.8\sigma$ instead in our formula to obtain agreement of $\sim 10$\% as in the case of peaks.  The issue of how to  best weight halos (or halo tracers) has recently started to be explored in the literature.  So far most efforts has been devoted to reduce stochasticity  and therefore improve the signal-to-noise ratio in power spectrum measurements \citep{2009PhRvL.103i1303S,2010PhRvD..82d3515H,2012PhRvD..86j3513H}. For example  \cite{Cai/Bernstein/Sheth:2011} argue that  an optimal weighting scheme which is a mix of bias weighting and mass weighting is best for reducing stochasticity. Different weighting schemes might be needed depending on the goal: reducing stochasticity, obtaining a shape closer to the linear power spectrum or closer to available  theoretical predictions.  We leave improvements and further developments in this direction to future work.

\section{Effect on the BAO feature}
\label{sec:BAO}
We now  concentrate  on the scale-dependent bias effect around the BAO feature, which should not be  heavily affected by evolution bias or by the weighting scheme chosen. The BAO feature is affected by non-linearities, and one could consider this as a form of evolution bias. The effect of non-linearities on the BAO feature have been extensively studied and can be modelled to high accuracy (see  e.g., \cite{Eisenstein/etal:2007} and references therein). Despite the BAO signal being small, it has been measured exquisitely well (e.g., \citet{BOSSPaper}). It is therefore important to  investigate  the implications of the effects considered in this paper on the measurement of the location of the BAO feature both in terms of a possible bias (shift) or increase in errors. (Most of) all BAO measurements reported in the literature  are ``protected" for  variations in the broadband shape of the power spectrum (or correlation function)   via a marginalization procedure see. e.g., \cite{Seo/Eisenstein:2007} and references therein.

\begin{figure}
\centering
\includegraphics[width=0.48\columnwidth]{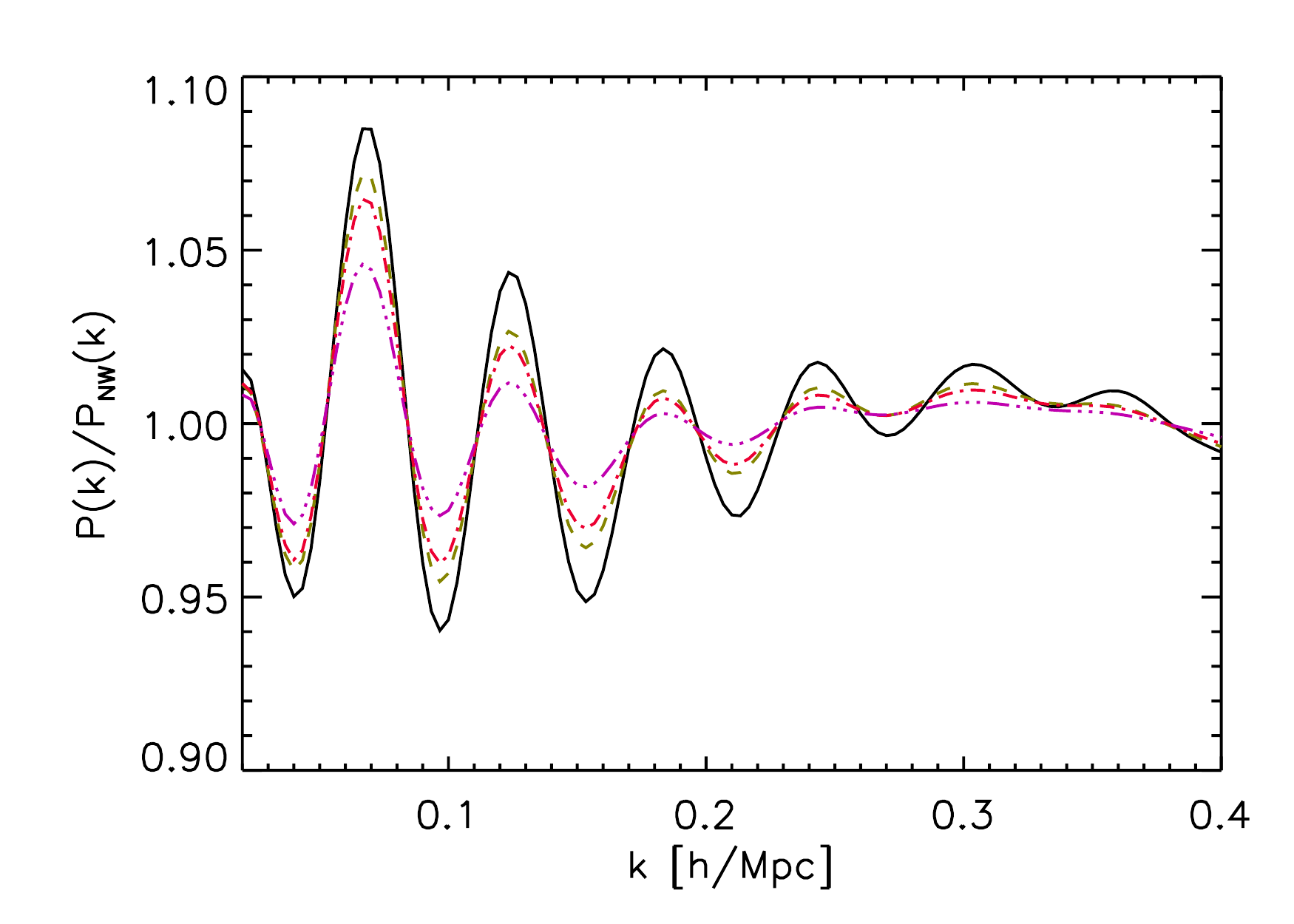}
\includegraphics[width=0.48\columnwidth]{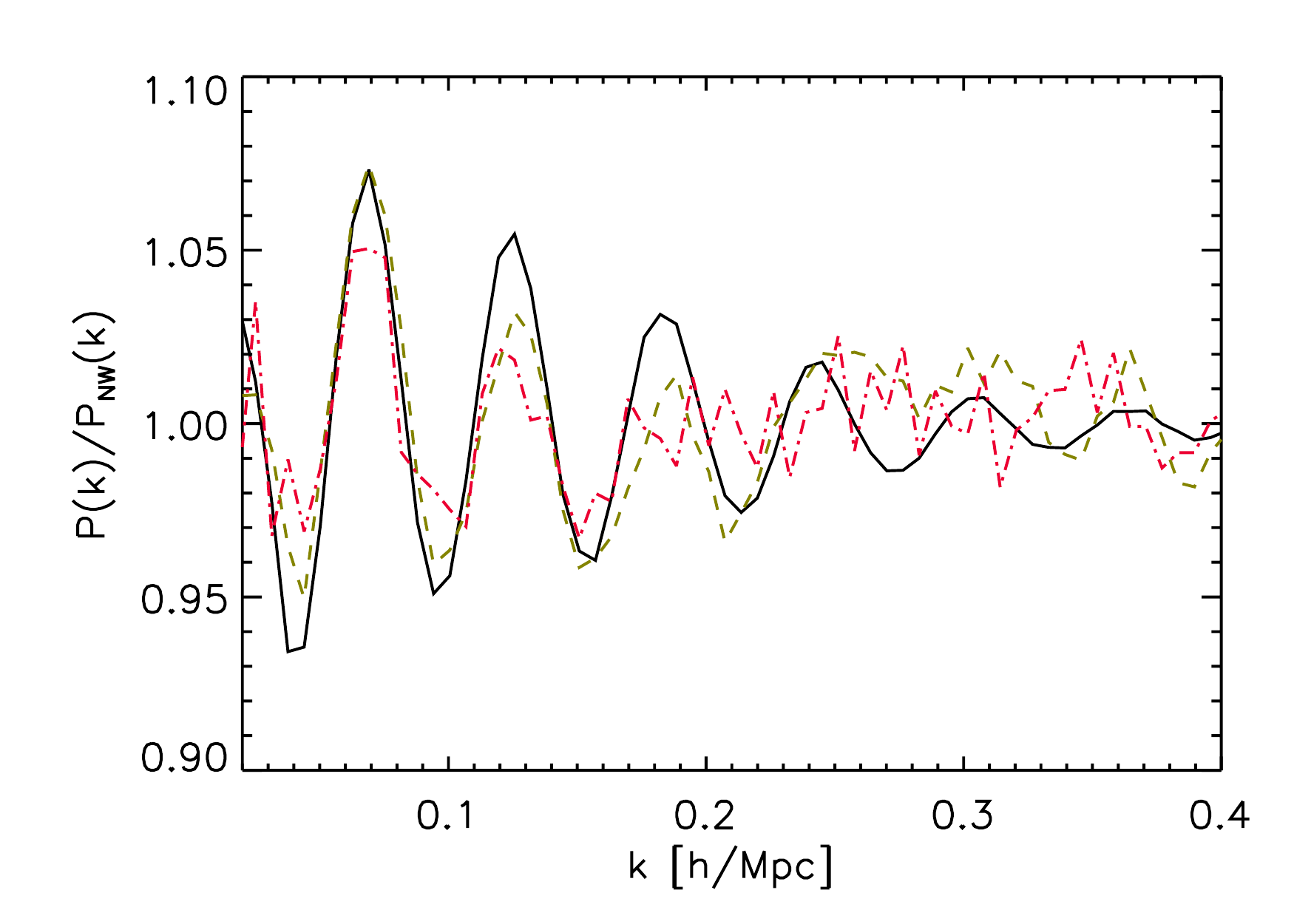}
\caption{The BAO feature as $P(k)/P_{\rm NW}$. The left panel shows  the theory prediction obtained by Fourier transforming eq.\ref{eq:corrana}, the right panel  shows the mean of the 20 Gaussian realisations. Solid/Black for matter, green/dashed for 2$\sigma$ peaks,   red/dot-dashed for 3$\sigma$ peaks; purple/dashed-dot-dot-dot for  4$\sigma$ peaks.}
\label{fig:baopk}
\end{figure}

In Fig.\ref{fig:baopk} we show  the BAO feature as the  ratio between the  measured power spectrum (be it dark matter or tracers)  and a smoothed version with no wiggles ($P_{\rm NW}$). The figure shows the underlying (dark matter) and peaks above different thresholds (see caption for details). This shows that the  localised scale-dependence of the bias does not move the BAO feature but reduces its amplitude. The smoothing is more marked for higher thresholds. 
 We find that the effects can be well described by the following:
 \begin{equation}
 P_{\rm pk}(k)=(P_{\rm DM}-P_{\rm NW})\exp\left(-\frac{k \Sigma_{\rm pk}^2}{2}\right)+P_{\rm NW}
 \end{equation}
 The parameter $\Sigma_{\rm pk}$ takes values of 2.5, 3 and 4.2 at 2,3, and 4 $\sigma$ thresholds respectively.
 Note that the effect  of non-linear evolution on the BAO can be described by a very similar expression where the argument of the exponential damping is $k^2 \Sigma_{\rm nl}^2/2$. The two effects have a different $k$ dependence; still we can compare their magnitude at $k=0.1$ $h$/Mpc: the bias effect for 2$\sigma$ peaks is equal to the non-linear effect at $z=0.3$ ($\Sigma_{\rm nl}\sim 8$) but becomes more important for higher thresholds.
 While it is reassuring that this effect does not change the location of the BAO feature it might have practical  implications.
 The signal-to-noise for measurements that depend on the BAO location is  usually computed adopting a model with a fixed BAO smoothing parameter  $\Sigma_{\rm nl}$: should the  chosen $\Sigma_{\rm nl}$ be smaller than the effective one (which would be  a combination of  $\Sigma_{\rm nl}$ and  $\Sigma_{\rm pk}$) then the signal-to-noise would be overestimated. 
 
There are consequences also for survey selection considerations: highly biased tracers are used to beat shot noise --in technical terms, to maximise $nP$ where $n$ denotes the tracer number density--. However highly biased tracers will have a reduced BAO feature: it may be advantageous to select less rare and less biased tracers if they carry a more pronounced signal. This will be presented elsewhere.
 
\section{Summary and Conclusions}
\label{Conclusions}
The bias of dark matter tracers (be it galaxies or, more simple objects like  halos or  density peaks) is a very complicated, non-linear, non-local function that relates the density of tracers to the density of dark matter. A lot of effort is going into understanding and modelling the bias. Here we approached a simpler problem: that of modelling the  correlation properties of tracers. In particular we  present an analytic  expression for the  (N-point) correlation function of extrema in random gaussian fields, weighted by $1/|\det w|$.  The results are valid in any number of dimensions, but we focus on the two-point function in three dimensions, which is of most practical relevance.  Our main result is  thus Eq.~\ref{eq:corrana}.
In order to be able to arrive at a fully analytic result  we had to assume that observations could be suitably weighted. 

Since most extrema above practically all interesting thresholds ($> 2\sigma$) are peaks, we find that one can identify the clustering of extrema with the clustering of peaks. 
Because, for a high enough threshold, dark matter halos correspond to peaks of the initial field (and vice versa),  we argue that this provides also an analytic description for the clustering of dark matter halos.

On the other hand the clustering properties of peaks may be of interest by themselves, for observations that produce directly density maps (e.g., weak lensing).

We find that the presence of non-zero derivatives in the underlying power spectrum introduce scale-dependent features in the bias which would otherwise be constant in peak theory. We identify two scale-dependent features, one broad-band which is  most affected by the  choice of weighting and evolution bias and a localised one, which is expected to be robust to these effects.
The localised, scale-dependent feature in the bias coincides with the location of the  baryon acoustic feature (BAO). Its effect is  to  smooth the BAO feature but, conveniently, it does not move it and we provide a simple analytic formula to describe it.

We have tested that the analytic formula we present describes accurately the clustering properties of peaks in a suite of Gaussian realisations. 

We find that the evolution bias appears to be relatively small, in other words clustering properties of peaks in the  low redshift, highly non-linear field are very similar to those of the high-redshift Gaussian field. Given the fact that halos are identified with density peaks, this opens up the possibility to use our findings to describe halo clustering. We have started exploring this possibility although to get a detailed, quantitative  description the choice of the  weighting scheme used is crucial. In addition for halos  the correspondence  between  actual threshold and  the theoretical one might not be straightforward.  We have explored  the performance of out initial ansatz for an inverse halo mass halo weighting scheme, which we find encouraging.

\section*{Acknowledgments}
LV  and FRGS are supported by the European Research Council under the European Community's Seventh Framework Programme FP7-IDEAS-Phys.LSS 240117. LV and RJ acknowledge support of  Mineco grant FPA2011-29678- C02-02.

\section*{Appendix A: Realistic weighting scheme ``ansatz''}
\begin{figure}
\centering
\includegraphics[width=0.5\columnwidth]{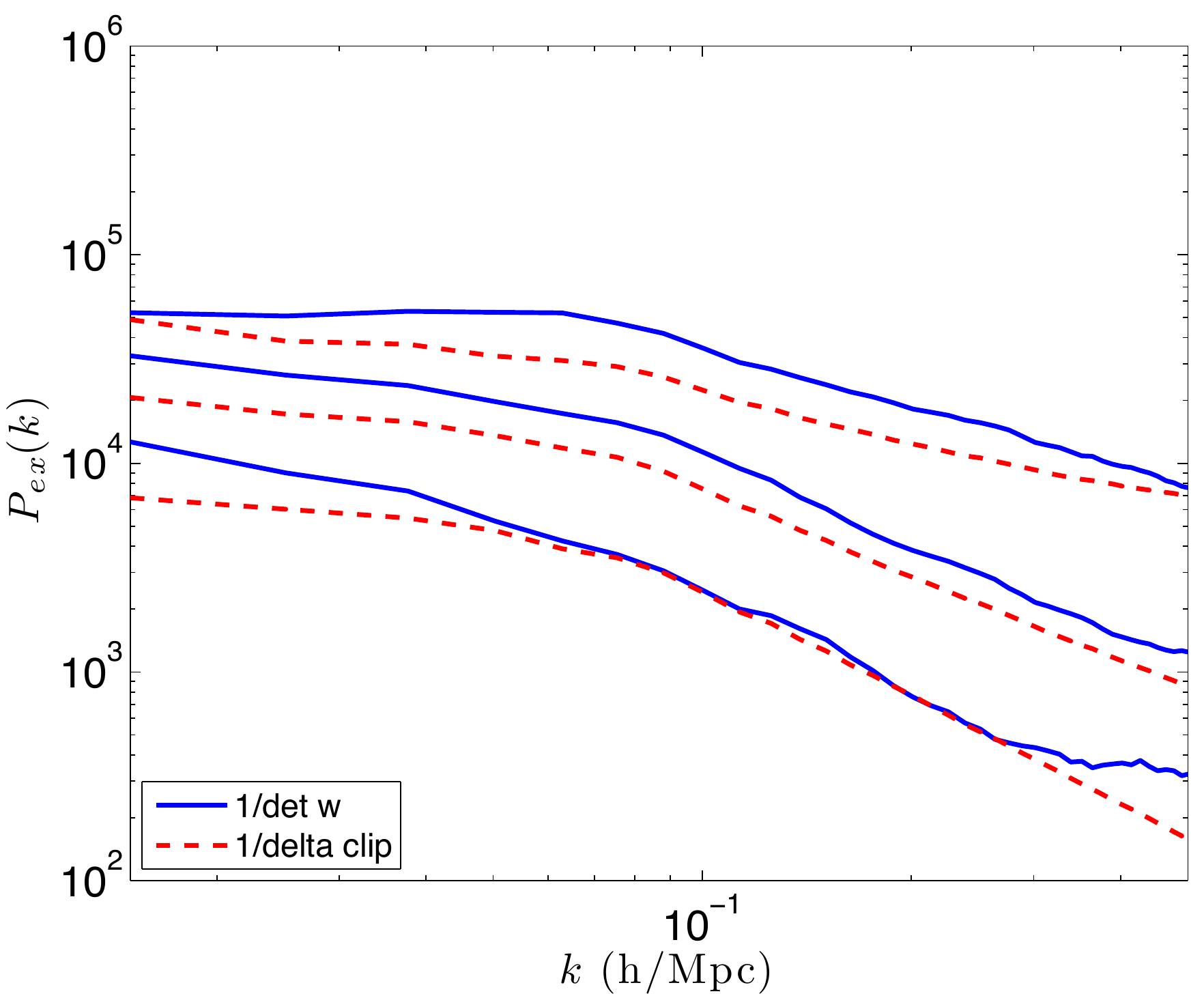}
\caption{The power spectra  of the  z=127 snapshot of the millennium simulation, for two different weighting schemes, for the solid line  we apply a weighting of $1/|\det w|$ to all peaks above $1,2$ and $3 \sigma$, the dashed line corresponds to weighting the whole volume above the threshold by $1/\delta$. The upturn in power on small scales appears to be a consequence of the peaks being distributed in a non Poissonian manner. }\label{fig:detw}
\end{figure}

The $1/|\det w|$ weighting scheme explored in this work is not straightforward to implement with real data, therefore here  we attempt to seek a simpler and more practical weighting scheme which may act as a suitable proxy.  The second derivative of the density field is a difficult quantity to extract directly from observations, however the curvature of a given peak is correlated with its amplitude: this is our motivation for the following ``ansatz", an inverse weighting scheme, corresponding to a local density transformation which penalises the higher density regions. 

 \begin{eqnarray}
\delta_{\rm inv} (x)  &=&  \frac{1}{\delta(x)} \, \, \,  (\delta (x) > \delta_0 ) \\
\delta_{\rm inv} (x)  &=&  0           \quad \, \, \,   (\delta (x) \leq  \delta_0) 
\end{eqnarray}

 The power spectra resulting from this weighting scheme, and that of $1/| \det w|$, can be seen in Figure \ref{fig:detw}, for thresholds of $1$, $2$, and $3 \sigma$.
 This  ``ansatz"  is probably a good starting point, but clearly a better approximation is needed.
 When considering halos,  we look for a weight that depends on halo mass (which is a quantity easier to estimate than  $| \det w|$ or even $\delta$). The $1/|\det w|$ weighting  down-weights high narrow peaks, we therefore propose as a first   ansatz an inverse halo mass weighting. Further investigation along these lines is left for future work.

\end{document}